%% file: ms_v3.tex
\documentclass[12pt,preprint]{aastex}

\slugcomment{Edit 7 Dec 2008}

\shorttitle{The Magnetic field in G5.89} \shortauthors{Tang et
al.}

\begin{document}


\title{Evolution of Magnetic Fields in High Mass Star Formation:
SMA dust polarization image of the UCHII region G5.89-0.39}


\author{Ya-Wen Tang}
\affil{Physics Department, National Taiwan University \&
\\ Institute of Astronomy and Astrophysics, Academia Sinica,
Taiwan, R.O.C.}

\author{Paul T. P. Ho}
\affil{Institute of Astronomy and Astrophysics, Academia Sinica,
Taiwan, R.O.C. \& \\ Harvard-Smithsonian Center for Astrophysics,
U.S.A.}

\author{Josep Miquel Girart}
\affil{Institut de Ci\`{e}ncies de l' Espai (CSIC-IEEC),
Catalonia, Spain}

\author{Ramprasad Rao}
\affil{Institute of Astronomy and Astrophysics, Academia Sinica,
Taiwan, R.O.C.}

\author{Patrick Koch}
\affil{Institute of Astronomy and Astrophysics, Academia Sinica,
Taiwan, R.O.C.}

\author{Shih-Ping Lai}
\affil{Institute of Astronomy and Department of Physics, National
Tsing Hua University \& \\Institute of Astronomy and Astrophysics,
Academia Sinica, Taiwan, R.O.C.}



\begin{abstract}

We report high angular resolution (3$\arcsec$) Submillimeter Array
(SMA) observations of the molecular cloud associated with the
Ultra-Compact HII region G5.89-0.39. Imaged dust continuum
emission at 870$\mu$m reveals significant linear polarization. The
position angles (PAs) of the polarization vary enormously but
smoothly in a region of 2$\times$10$^{4}$ AU. Based on the
distribution of the PAs and the associated structures, the
polarized emission can be separated roughly into two components.
The component "x" is associated with a well defined dust ridge at
870 $\mu$m, and is likely tracing a compressed B field. The
component "o" is located at the periphery of the dust ridge and is
probably from the original B field associated with a pre-existing
extended structure. The global B field morphology in G5.89, as
inferred from the PAs, is clearly disturbed by the expansion of
the HII region and the molecular outflows.  Using the
Chandrasekhar-Fermi method, we estimate from the smoothness of the
field structures that the B field strength in the plane of sky can
be no more than 2$-$3 mG. We then compare the energy densities in
the radiation, the B field, and the mechanical motions as deduced
from the C$^{17}$O 3-2 line emission.  We conclude that the B
field structures are already overwhelmed and dominated by the
radiation, outflows, and turbulence from the newly formed massive
stars.
\end{abstract}



\section{Introduction}

One of the main puzzles in the study of star formation is the low
star formation efficiency in molecular clouds. Since molecular
clouds are known to be cold, the thermal pressure is small. Hence,
if there are no other supporting forces against gravity, the
free-fall time scale will be short and the star formation rate
will be much higher than what is observed. Magnetic (B) fields
have been suggested to play the primary role in providing a
supporting force to slow down the collapsing process (see the
reviews by Shu et al. (1999) and Mouschovias \& Ciolek (1999)). In
these models, the B field is strong enough and has an orderly
structure in the molecular cloud. The B field lines, which are
anchored to the ionized particles, will then be dragged in along
the direction of accretion, only when the ambipolar diffusion
process allows the neutral component to slip pass the ionized
component. In the standard low-mass star formation model (Galli \&
Shu 1993; Fiedler \& Mouschovias 1993), an hourglass-like B field
morphology is expected with an accreting disk near the center of
the pinched field. Alternatively, turbulence has also been
suggested as a viable source of support against contraction (see
reviews by Mac Low \& Klessen (2004) and Elmegreen \& Scalo
(2004)). The relative importance of B field and turbulence
continues to be a hot topic as the two methods of support will
lead to different scenarios for the star formation process.

Compared with the low mass stars, the formation process of high
mass stars is really poorly understood. High mass star forming regions,
because of their rarity, are usually at larger distances and are
always located in dense and massive regions, because they are
typically formed in a group. Hence, both poor resolution and
complexity have hampered past observational studies. Furthermore,
the environments of high mass star forming regions are very
different from the low mass case because of higher radiation
intensity, higher temperature, and stronger gravitational fields.
Will the B fields in massive star forming sites have a similar
morphology to the low mass cases?

Polarized emission from dust grains can be used to study the B
field in dense regions, because the dust grains are not spherical
in shape. They are thought to be aligned with their minor axes
parallel to the B field in most of the cases, even if the
alignment is not magnetic (Lazarian 2007). Due to the differences
in the emitted light perpendicular and parallel to the direction
of alignment, the observed thermal dust emission will be
polarized, the direction of polarization is then perpendicular to
the B field. Although the alignment mechanism of the dust grains
has been a difficult topic for decades (see review by Lazarian
2007), the radiation torques seem to be a promising mechanism to
align the dust grains with the B field (e.g., Draine \&
Weingartner 1996; Lazarian \& Hoang 2007). However, other
processes such as mechanical alignments by outflows can also be
important.

Polarized dust emission has been detected successfully at
arcsecond scales. The best example might be the low mass star
forming region NGC 1333 IRAS 4A (Girart, Rao \& Marrone 2006),
which reveals the classic predicted hourglass B field morphology.
Results on the massive star formation regions, such as W51 e1/e2
cores (Lai et al. 2001), NGC2024 FIR5 (Lai et al. 2002), DR21 (OH)
(Lai et al. 2003), G30.79 FIR 10 (Cortes et al. 2006) and
G34.4+0.23 MM (Cortes et al. 2008), typically show an organized
and smooth B field morphology.  However, this could be due to a
lack of spatial resolution. Indeed, for the nearby high mass cases
such as Orion KL (Rao et al. 1998) and NGC2071IR (Cortes,
Crutcher, \& Matthews 2006), abrupt changes of the polarization
direction on small physical scales have been seen, which may
suggest mechanical alignments by outflows as proposed by these
authors. Whether high mass star forming regions will all show
complicated B field structures on small scales remains to be
examined.

In this study, we report on one of the first SMA
measurements of dust polarization for a high mass star forming region,
G5.89-0.39
(hereafter, G5.89). The linearly polarized thermal dust emission
is used to map the B field at $\sim$3$\arcsec$ resolution, and
the C$^{17}$O 3-2 line is used to study the structure and
kinematics of the dense molecular cloud. The description of the
source, the observations and the data analysis, the results, and
the discussion are in Sec. 2, 3, 4 and 5, respectively. The
conclusions and summary are in Sec. 6.

\section{Source Description}
G5.89 is a shell-like ultracompact HII (UCHII) region (Wood et al.
1989) at a distance of 2 kpc (Acord et al. 1998). The UCHII region
is 0.01 pc in size, and its dynamical age is 600 years, estimated
from the expansion velocity (Acord et al. 1998). Observations of
the K$_{s}$ and $L'$ magnitudes and color by Feldt et al. (2003)
suggest that G5.89 contains an O5 V star.

Just as in other cases of massive stars, G5.89 contains most
likely a cluster of stars. The detections of associated H$_{2}$O
masers (Hofner \& Churchwell 1996), OH masers (Stark et al. 2007;
Fish et al. 2005) and class I CH$_{3}$OH masers (Kurtz et al.
2004), suggest that multiple stars have formed in this region.
Furthermore, the morphology of the detected molecular outflows
also suggest the presence of multiple driving sources, because
different orientations are observed in different tracers. In CO
1-0, the large scale outflow is almost in the east-west direction
(Harvey \& Forveille 1988; Watson et al. 2007). In C$^{34}$S and
the OH masers, the outflow is in the north-south direction
(Cesaroni et al. 1991; Zijlstra et al. 1990). In SiO 5$-$4, the
outflow is at a position angle (PA) of 28$\degr$ (Sollins et al.
2004).  In the CO 3-2 line, the outflows (Hunter et al. 2008) are
in the north-south direction and at the PA of 131$\degr$, and the
latter one is associated with the Br$\gamma$ outflow (Puga et al.
2006). In addition, the detected 870 $\mu$m emission has also been
resolved into multiple peaks (labelled in Fig. 1(a); Hunter et al.
2008). The different masers, the multiple outflows, and the
multiple dust peaks, are all consistent with the formation of a
cluster of young stars.

G5.89 should be expected to have a substantial impact on its
environment. In terms of the total energy in outflows in this
region, G5.89 is definitely one of the most powerful groups of
outflows ever detected (Churchwell 1997).


\section{Observation and Data Analysis}
The observations were carried out on July 27, 2006 and September
10, 2006 using the Submillimeter Array (Ho, Moran \& Lo
(2004))\footnote{The Submillimeter Array is a joint project
between the Smithsonian Astrophysical Observatory and the Academia
Sinica Institute of Astronomy and Astrophysics and is funded by
the Smithsonian Institution and the Academia Sinica.} in the
compact configuration, with 7 of the 8 antennas available for both
tracks.
The projected lengths of the baselines ranged from 6.5 to 70
k$\lambda$ ($\lambda\approx$870$\mu$m). Therefore, our
observational results are insensitive to structures
larger than 39$\arcsec$.
The SMA receivers are intrinsically
linearly polarized and only one polarization is available at the
current time. Thus, quarter-wave plates (see Marrone \& Rao 2008)
were installed in order to convert the linear polarization (LP) to
circular polarization (CP). The quarter-wave plates were rotated
by 90$ \degr$ on a 5 minutes cycle using a Walsh function to
switch between 16 steps in order to sample all the 4 Stokes
parameters. The integration time spent on the source in each step
was approximately 15 seconds. The overhead required in switching
between the different states was approximately 5 seconds. In each
cycle all four cross-correlations (LL, LR, RL, and RR) were each
calculated 4 times. The data were then averaged over this complete
cycle in order to obtain quasi simultaneous dual polarization
visibilities. We assume that the smearing due to the change of the
polarization angles on this time scale is negligible.

The local oscillator frequency was tuned to 341.482 GHz. With a 2
GHz bandwidth in each sideband we were able to cover the frequency
range from 345.5 to 347.5 GHz and from 335.5 to 337.5 GHz in the
upper and lower sideband, respectively. The correlator was set to
a uniform frequency resolution of 0.65 MHz ($\sim$ 0.7 km
s$^{-1}$) for both sidebands. While our main emphasis was to map
the polarized continuum emission from the dust, we were also able
to detect a number of  molecular lines simultaneously. These
results will be published separately.

Generally, the conversions of the LP to CP of the receivers are
not perfect. This non-ideal characteristic of the receiver will
cause an unpolarized source to appear polarized, which is known as
instrumental polarization or leakage. Nevertheless, these leakage
terms (see Sault, Hamaker, \& Bregman 1996) can be calibrated by
observing a strong linearly polarized quasar. In this study, the
leakage and bandpass were calibrated by observing 3c279 for the
first track and 3c454.3 for the second track. Both sources were
observed for 2 hours while they were transiting in order to get
the best coverage of parallactic angles. The leakage terms are
frequency dependent, $\sim$1\% and $\sim$3\% for the upper and
lower sideband before the calibration, respectively. After
calibration, the leakage is less than 0.5$\%$ in both sidebands.
Besides the calibration for the polarization leakage, the
amplitudes and phases
were calibrated by observing the quasars 1626-298 and 1924-292
every 18 minutes. These two gain calibrators in both tracks were
used because of the availabilities of the calibrators during the
observations. Finally, the absolute flux scale was calibrated
using Callisto.

The data were calibrated and analyzed using the MIRIAD package
(Sault, Teuben, \& Wright 1995). After the standard gain
calibration, self-calibration was also performed by selecting the
visibilities of G5.89 with uv distances longer than 30 k$\lambda$.
As a result, the sidelobes and the noise level of the Stokes $I$
image were reduced by a factor of 2. In order to get the images from
the measured visibilities, the task INVERT in the MIRIAD package
was used. The Stokes $Q$ and $U$ maps are crucial for the
derivation of the polarization segments. We used the dirty maps of
$Q$ and $U$ to derive the polarization to avoid a possible bias
introduced from the CLEAN process. The Stokes $I$ map shown in
this paper is after CLEAN.

The Stokes $I$, $Q$ and $U$ images of the continuum were
constructed with natural weighting in order to get a better S/N
ratio for the polarization. The final synthesized beam is
$3\arcsec.0 \times1\arcsec.9$ with the natural weighting. The
C$^{17}$O images are presented with a robust weighting 0.5 in
order to get a higher angular resolution, and the synthesized beam
is 2.8$\arcsec \times$1.8$\arcsec$ with PA of 13$\degr$. The noise
levels of the $I$, $Q$ and $U$ images are $\sim$ 30, 5 and 5 mJy
Beam$^{-1}$, respectively. Note that the noise level of the Stokes
$I$ image is much larger than the ones in the Stokes $Q$ and $U$
images. The large noise level of the $I$ image is most likely due
to the extended structure, which can not be recovered with our
limited and incomplete uv sampling. The strength (I$_{p}$) and
percentage ($P$) of the linearly polarized emission is calculated
from: $I_{p}^{2} = Q^2 +U^2-\sigma_{Q,U}^{2}$ and $P$ = I$_{p}$/I,
respectively. The term $\sigma_{Q,U}$ is the noise level of the
Stokes $Q$ and $U$ images, and it is the bias correction due to
the positive measure of I$_{p}$. The noise of I$_{p}$
($\sigma_{I_p}$) is thus 5 mJy Beam$^{-1}$. The presented
polarization is derived using the task IMPOL in the MIRIAD
package, where the bias correction of $\sigma_{I_p}$ is included.

\section{Results}
In this section, we present the observational results of the dust
continuum and the dust polarization at 870$\mu$m, and the
C$^{17}$O 3-2 emission line. No polarization was detected in the
CO 3-2 emission line.

\subsection{Continuum Emission}
The total continuum emission at 870 $\mu$m, shown in Fig. 1(a), is
resolved with a total integrated flux density 12.6$\pm$1.3 Jy. In
general, the morphology of the continuum emission at 870$\mu$m is
similar to the emission at 1.3 mm by Sollins et al. (2004).
However, the 870$\mu$m emission peaks at $\sim$ 1$\arcsec$ west of
the position of the O5 star, which is offset toward the north-west by
$\sim$1$\arcsec$.7 from the peak of the 1.3 mm continuum emission.
Because there is still a significant contribution from the
free-free emission to the continuum at 870 $\mu$m and at 1.3mm,
the differences between the 870 $\mu$m and 1.3 mm maps most likely
result from the increasing contribution from the dust emission as
compared to the free-free emission at shorter wavelengths. Due to
the importance of a correct dust continuum image in the derivation
of the polarization, we describe here how the free-free continuum
was estimated and removed from the 870$\mu$m total continuum emission.

\subsubsection{Removing the free-free emission}
The free-free continuum at 2cm (shown in color scale in Fig.1 (a)
and (b)) was imaged from the VLA archival database observed on
August 7, 1986. The VLA synthesized beam of the 2cm free-free
image is 0$ \arcsec.92\times$0$\arcsec$.45 with natural weighting
of the uv data. Since the free-free shell is expanding at a rate
of 2.5 mas year$^{-1}$ (Acord et al. 1998), at a distance of 2
kpc, this expansion motion over the intervening 20 years is
negligible within the synthesized beam of our SMA observation.

The contribution from the free-free continuum was removed by the
following steps. Firstly, we adopted a spectral index
$\alpha=-$0.154 calculated in Hunter et al. (2008) for the
free-free continuum emission between 2cm to 870$\mu$m. The
resulting estimated free-free continuum strength at 870$\mu$m was
4.9 Jy. Secondly, we further assumed that the morphology of the
free-free continuum at 870 $\mu$m and at 2cm were identical. We
then smoothed the VLA 2cm image to the SMA resolution and scaled
the total flux density to 4.9 Jy. Finally, we subtracted this
image from the total continuum at 870 $\mu$m. The resultant 870
$\mu$m dust continuum image is shown in Fig. 1(b). The total flux
density of the dust continuum is therefore 7.7$\pm$0.8 Jy.

\subsubsection{Dust continuum: mass and morphology}
The corresponding gas mass (M$_{gas}$) was calculated from the
flux density of the dust continuum at 870 $\mu$m following Lis et
al. (1998):

\begin{equation}\label{mh2}
    M_{gas} = \frac{2\lambda^{3} Ra \rho d^{2}}{3hcQ(\lambda)J(\lambda,T_d)}S(\lambda)
\end{equation}

Here, we assumed a gas-to-dust mass ratio $R$ of 100, a grain
radius $a$ of 0.1 $\mu$m, a mean grain mass density $\rho$ of 3 g
cm$^{-3}$, a distance to the source $d$ of 2 kpc, a dust
temperature $T_{d}$ of 44 K, an observed flux density S$(\lambda)$
of 7.7 Jy, the Planck factor $J(\lambda, T_{d})=[exp(hc/\lambda k
T_{d})-1]^{-1}$. $h$, $c$ and $k$ are the Planck constant, the
speed of light and the Boltzmann constant, respectively. The grain
emissivity $Q$($\lambda$) was estimated to be $1.5\times 10^{-5}$
after assuming $Q(350\mu m)$ of $7.5\times 10^{-4}$ and $\beta$ of
2 (cold dust component), and using the relation
$Q(\lambda)=Q(350\mu m)(350\mu m/\lambda)^{\beta}$ (Hunter et al.
2000). As suggested in the same paper, the dust emission can be
modeled by two temperature components, with the emission dominated
by the colder component at T$_{d}$ $\sim$ 44 K. We adopted this
value for T$_{d}$, and therefore, the mass given here refers only
to the cold component and is an underestimate of the total mass.
The derived gas mass of the dust core M$_{gas}$ is $\sim$ 300
M$_{\sun}$, with a number density $n_{H_{2}}=$ 5.3$\times$10$^{6}$
cm$^{-3}$ averaged over the emission region. The sizescale along
the line of sight is assumed to be 0.13 pc, which is the diameter
of the circle with the equivalent emission area.

The dust emission presented in Fig. 1(b) has an extension toward
the northeast, east and southwest and has a steep roll off on the
northwestern edge of the ridge. In the higher angular resolution
(0.8$\arcsec$) observation at the same wavelength by Hunter et al.
(2008), the dust core is resolved into 5 peaks, where the two
strongest peaks align in the north-south direction to the west of
the O5 star. The dust continuum emission associated with SMA-N,
SMA-1 and SMA-2 is called \textit{sharp dust ridge} hereafter
because of its strong emission and its morphology. There is no
peak detected at the position of the O5 star. It is likely that
the O5 star is located in a dust-free cavity, as proposed by Feldt
et al. (1999) and Hunter et al. (2008).

\subsection{Dust polarization}
We first compare the dust polarization derived from the 870 $\mu$m
total continuum (Fig. 1(c)) and from the 870 $\mu$m dust continuum
(Fig. 1(d)). In both cases the derived polarization is at the same
location with the same PAs. The only difference of the
polarization in Fig. 1(c) and 1(d) is that the percentage of
polarization near the HII region is increased in Fig. 1(d). This
is because of the fact that the free-free continuum is not
polarized, and the $Q$ and $U$ components are not affected by the
free-free continuum subtraction. Therefore, the expected
polarization percentage will increase when the free-free continuum
is removed from the 870 $\mu$m continuum. The total detected
polarized intensity I$_{p}$ is 59 mJy. All the polarization shown
in the figures besides Fig. 1(c) is calculated from the derived
dust continuum image. The off-set positions, percentages and PAs
of the polarization segments are listed in Table 1.

\subsubsection{Morphology of the detected polarization}

The polarized emission is not uniformly distributed. Detected
polarization at
2$\sigma_{I_p}$ are shown as blue segments and detections above
3$\sigma_{I_p}$ are shown by red
segments.  Most of the polarized emission is
located in the northern half of the dust core close to the HII
region and appears as 4 patches, mostly with $\sigma_{I_p} \geq 3$
(Fig. 2(a) in color scale). There is a sharp gap where no
polarization is detected extending from the NE to the SW across
the O star. The southern half of the dust core is free of
polarization, except for a few positions at the edge of the dust
core. However, the polarization in the south half of the dust core
is at 2 to 3$\sigma_{I_p}$ level only.
We will focus our discussions on the more significant detections
in the core of the cloud.

We separate the polarized emission into two groups.  We are guided
principally by the fact that one group is associated with the periphery
of the total dust emission, while the other group tracks the strongest
parts of the total dust emission.
The polarized patches to the east of the O star and to the west
of the Br$\gamma$ outflow source have similar PAs of $\sim 50
\degr$ (Fig. 2(b)). These polarization segments are located at the
fainter edges of the higher resolution 870 $\mu$m dust continuum
image (Fig. 2(c); Hunter et al. 2008) and at the less steep part
of the 3$\arcsec$ resolution image (this paper). This may suggest
that this polarization originates from a more extended overall
structure, rather than from the detected condensations. Therefore,
these polarization segments are suggested to be the component "o"
(defined in the next section). The rest of the polarization in the
northern part is all next to the sharp gap where no polarization
is detected. Most of the polarization is on the 870$\mu$m
\textit{sharp dust ridge} observed with 0.8$\arcsec$ resolution,
except for the ones at the NE and SW ends where the polarization
patches stretch toward the extended structure. At these NE and SW
ends the polarization is probably the sum of the extended and the
condensed structures. These polarization segments are suggested to
belong to the component "x".

The 0.8$\arcsec$ resolution observations show that there is a hole
in the southern part of the detected dust continuum. This hole is
not resolved with the 3$\arcsec$ synthesized beam of our map. That
may explain why polarization is not detected at this position.
Here, and also for the dust ridge sharply defined with
0.8$\arcsec$ resolution, the dust polarization is sensitive to the
underlying structures and can help to identify unresolved features
which are smaller than our resolution.

\subsubsection{Distribution of the polarization segments}
The detected PAs vary enormously over the entire map, ranging from
$-60\degr$ to 61$\degr$ (Fig. 3(a)). Nevertheless, they vary
smoothly along the dust ridge and show organized patches. We have
roughly separated the polarized emission into two different
components according to their locations (as discussed in Sec.
4.2.1) and their PAs. The "o" component is probably from an
extended structure with PAs ranging from 33$\degr$ to 61 $\degr$.
The mean PA weighted with the observational uncertainties of
component "o" is 49$\pm$3$\degr$, with a standard deviation of
11$\degr$. The "x" component associated with the \textit{sharp
dust ridge} has PAs ranging from $-60\degr$ to 4$\degr$. Its
weighted mean PA is $-$24$\pm$1$\degr$, with a standard deviation
of 18$\degr$. If the polarization were not separated into two
components, the weighted mean PA is $-$9$\degr$ with a standard
deviation of 39$\degr$.

The relation between the percentage of polarization and the intensity
is shown in Fig. 3(b). The percentage of polarization decreases
towards the denser regions, which has already been seen for
other star formation sites, such as the ones listed in Sec.
1. This is possibly due to a decreasing alignment efficiency in high
density regions, because the radiation torques are relatively
ineffective (Lazarian \& Hoang 2007). It can also be due to the
geometrical effects, such as differences in the viewing angles
(Gon\c{c}alves et al. 2005), or due to the results from averaging
over a
more complicated underlying field morphology.

\subsection{C$^{17}$O 3-2 emission line}
In order to trace the physical environments and the gas kinematics
in G5.89, we choose to use the C$^{17}$O 3-2 emission line because
of its relatively simple chemistry. The critical density of
C$^{17}$O 3-2 is $\sim$ 10$^{5}$ (cm$^{-3}$), assuming a
cross-section of 10$^{-16}$ (cm$^{-2}$) and a velocity of 1 km
s$^{-1}$, and therefore, it will trace both the relative lower
(n$_{H_2}$ $\sim$10$^{5}$ (cm$^{-3}$)) and higher (n$_{H_2}$$\sim$
10$^{6}$ (cm$^{-3}$)) density regions. Although its critical
density is much smaller than the estimated gas density of
5.3$\times$10$^{6}$ (cm$^{-3}$) from the dust continuum, it is
apparently tracing the same regions as the dust continuum because
of the similar morphology of the integrated intensity image, shown
in the next section. We therefore assume that the kinematics
traced by C$^{17}$O represents the bulk majority of the molecular
cloud and that it is well correlated with the dust continuum.

\subsubsection{Morphology of C$^{17}$O 3-2 emission}
The emission of the C$^{17}$O 3-2 line covers a large velocity
range, from $-$7 to 28 km s$^{-1}$, as shown in the channel maps
in Fig. 4. The majority of the gas traced by the C$^{17}$O 3-2
line is relatively quiescent and has a morphology similar to the
870$\mu$m dust continuum emission. Besides the components which
trace the dust continuum, an arc feature is seen in the south-east
corner of the panel covering 10 to 15 km s$^{-1}$. There is no associated
870$\mu$m dust continuum detected at this location, probably due
to the low total column density or mass of this feature. Another
feature seen in the more quiescent gas is the clump extending towards the
south of the dust core (see the panel covering 6 to
10 km s$^{-1}$ in Fig. 4). This clump has a similar morphology as
seen in the 870 $\mu$m dust continuum where no polarization has
been found. At the higher velocity ends, i.e. from $-$7 to $-$3 km
s$^{-1}$ and from 23 to 28 km s$^{-1}$, the emission appears at
the 870$\mu$m dust ridge. This suggests that at the \textit{sharp
dust ridge}, there are high velocity components besides the
majority of quiescent material. Furthermore, the brightest HII
features appear correlated with the strongest C$^{17}$O emission,
especially at low velocities (v$_{lsr}=$ 6 to 15 km s$^{-1}$),
which may point toward an interaction between the molecular gas
and the HII region.

The total integrated intensity (0th moment) image (Fig.
5(\textit{upper-panel})) of the C$^{17}$O 3-2 emission line shows
a similar morphology as the 870 $\mu$m dust continuum. The
morphology of the C$^{17}$O gas to the west of the O star is
similar to the dense dust ridge, i.e. there is an extension from
north to south. The steep roll off of the dust continuum in the
north-west and an extension from NE to the west of the O star are
also seen in C$^{17}$O. Besides these similar features to the dust
continuum, a strong C$^{17}$O peak is found at position A, where
no dust continuum peak is detected. This feature A likely does
not have much mass, and we will not discuss its properties further
in this paper.

\subsubsection{Total gas mass from C$^{17}$O 3-2 line}
The total gas mass M$_{gas}$ in this region can be derived from
the C$^{17}$O 3-2 line. This provides a complementary estimate,
which is independent from the mass derived from the dust continuum
in Eq. 1. Assuming that the observed C$^{17}$O 3-2 line is
optically thin and in local thermal equilibrium (LTE), the mean
column density $N_{C^{17}O}$ is calculated following the standard
derivation of radiative transfer (see Rohlfs \& Wilson 2004):
\begin{equation}\label{1}
    N_{C^{17}O} = 1.3 \times 10^{13} \times
\frac{T_{R3-2}\triangle V}{D(n,T_{k})}
\end{equation}

Here, the T$_{R3-2}\triangle$V term is the mean flux density of
the entire emission region in K km s$^{-1}$. The D parameter
depends on the number density $n$ and the kinetic temperature
T$_{k}$ and is given by:
    \begin{displaymath}\label{1}
    D(n,T_{k})=f_{2}[J_{\nu}(T_{ex})-J_{\nu}(T_{bk})][1-exp(-16.597/T_{ex})],
    \end{displaymath}

where f$_2$ is the population fraction of C$^{17}$O molecules in
the J$=$2 state. T$_{ex}$ and T$_{bk}$ are the excitation and
background temperatures, respectively. The adopted value of D is
1.5 from the LVG calculation by Choi, Evans II $\&$ Jaffe (1993).
In their calculation, this D value is correct within a factor of 2
for 10 $<$ T$_{k}$$<$ 200 K in the LTE condition. The total gas
mass M$_{gas}$ is given by:

\begin{equation}\label{2}
    M_{gas} = \mu m_{H_{2}} d^2 \Omega \frac{N_{C^{17}O}}{X_{C^{17}O}}
\end{equation}

$\mu$ is 1.3, which is a correction factor for elements heavier
than hydrogen. m$_{H_2}$ is the mass of a hydrogen molecule. $d$
and $\Omega$ are the distance to the source and the solid angle of
the emission, respectively. The C$^{17}$O abundance $X_{C^{17}O}$
is assumed to be {\rm5 $\times$ 10$^{-8}$} (Frerking \& Langer
1982; Kramer et al. 1999). The derived mean $N_{C^{17}O}$ is {\rm
2$\times$10$^{16}$ cm$^{-2}$}. The mean gas number density
$n_{H_{2}}$ is {\rm 1.6$\times$10$^{6}$ $cm^{-3}$}, assuming the
size of the molecular cloud is 0.13 pc along the line of sight,
which is the diameter of the circle with the equivalent emission area.
The derived M$_{gas}$ from the C$^{17}$O 3-2 emission is $\sim$100
M$_{\sun}$.

The gas mass calculated using the C$^{17}$O 3-2 line is a factor
of 3 smaller than the value derived from the dust continuum (300
M$_{\sun}$). This difference has also been seen in the C$^{17}$O
survey towards the UCHII regions by Hofner et al. (2000). Their
M$_{gas}$ estimated from the measurement of the C$^{17}$O emission
tends to be a factor of 2 smaller than the measurement from the
dust continuum. The uncertainty of the estimate here possibly
results from the assumptions of the dust emissivity, the gas to
dust ratio, the abundance of the C$^{17}$O, and from the
possibility that C$^{17}$O might not be entirely optically thin.

\section{Discussion}
We discuss the possible reasons of the non-detected polarization
in the CO 3-2 line in the next paragraph. In order to interpret
our results, we have also analyzed the kinematics of the
molecular cloud in G5.89 using the C$^{17}$O 3-2 1st and 2nd
moment images, the position velocity (PV) diagrams, and the
spectra at various positions. The strength of the B field inferred
from the dust polarization is calculated using the
Chandrasekhar-Fermi method. A possible scenario of the dust
polarization is discussed based on the calculation of the mass to
flux ratio and the energy density.

\subsection{CO 3-2 polarization}

Under the presence of the B field, the molecular lines can be
linearly polarized if the molecules are immersed in an anisotropic
radiation field and the rate of radiative transitions is at least
comparable with the rate of collisional transitions. This effect
is called the Goldreich-Kylafis (G-K) effect (Goldreich \& Kylafis
(1981); Kylafis (1983)). The G-K effect provides a viable way to
probe the B field structure of the molecular cores, because the
polarization direction is either parallel or perpendicular to the
B field. The degree of the polarization depends on several
factors: the degree of anisotropy; the ratio of the collision rate
to the radiative rate; the optical depth of the line; and the
angle between the line of sight, the B field, and the axis of
symmetry of the velocity field.  In general, the maximum
polarization occurs when the line optical depth is $\sim$ 1
(Deguchi \& Watson 1984). Although the predicted polarization can
be as high as 10\%$-$20\%, the G-K effect is only detected in a
limited number of star formation sites: the molecular outflow as
traced by the CO molecular lines with BIMA in the source NGC 1333
IRAS 4A (Girart \& Crutcher 1999), and the outer low-density
envelope in G34.4+0.23 MM (Cortes et al. 2008), G30.79 FIR 10
(Cortes \& Crutcher 2006) and DR 21(OH) (Lai et al. 2003).  High
resolution observations are required to separate regions with
different physical conditions.

We have checked the polarization in the molecular lines. No
detection in the CO 3-2 and other emission lines was found. The
molecular outflows as seen in the CO 3-2 and SiO 8-7 emission
lines will be shown in Tang et al. (in prep.). We briefly discuss
the possible reasons for the lack of polarization in the molecular
lines here.

One possible reason is the high optical depth ($\tau$) of the CO
3-2 line. It has been shown by Goldreich and Kylafis (1981) that
the percentage of polarization depends on the value of $\tau$,
decreasing rapidly as the line becomes optically thick. When
corrected for multi-level populations, Deguchi \& Watson (1984)
suggested that the percentage of polarization decreases further by
about a factor of 2. In G5.89, $\tau$ of the CO 3$-$2 line is
$\sim$10 at v$_{lsr}$ = 25 km s$^{-1}$ (Choi et al. 1995), which
is the channel where the emission is strongest in our SMA
observation. Note that this emission does not peak at the
systematic velocity ($v_{sys}$) of 9.4 km s$^{-1}$, which is most
likely due to the missing extended structure which our observation
cannot reconstruct. We then estimate that the expected
percentage of polarization will be about 1.5$\%$, or a polarized flux
density of 0.5 Jy Beam$^{-1}$ for the CO 3-2 line, which is below
our sensitivity.

Besides an optimum $\tau$, the anisotropic physical conditions,
such as the velocity gradient and the density of the molecular
cloud, are needed to produce a polarized component from the
spectral line. The fraction and direction of polarization will
also change as a function of $\tau$ if there are external
radiation sources nearby (Cortes et al. 2005). Here, we are not
able to distinguish between these possible reasons.

\subsection{The kinematics traced by C$^{17}$O 3-2 emission line}
As shown in Sec. 4.3.1, high velocity components of the molecular
gas are traced by the C$^{17}$O 3-2 emission near the HII region.
Here, we examine the kinematics in G5.89.

The intensity weighted velocity (1st moment) image provides the
information on the line-of-sight motion (mean velocity). The
molecular cloud is red-shifted with respect to the v$_{sys}$ of
9.4 (km s$^{-1}$) in the NW (position $B$) and SE (position $D$)
of the O5 star (middle panel of Fig. 5). Next to the south of the
O5 star, a blue-shifted clump with respect to 9.4 km s$^{-1}$ is
detected. The molecular cloud in G5.89 has significant variations
in mean velocity within a radius of 5$\arcsec$ around the O5 star.

To further investigate the relative motions, the total velocity
dispersion $\delta v_{total}$ (2nd moment) image is also presented
(Fig. 5 (\textit{lower-panel})). $\delta v_{total}$ is related to
the spectral linewidth at full-width half maximum (FWHM) for a
Gaussian line profile: FWHM $=$ 2.355$\delta v_{total}$. Around
the HII region in G5.89, $\delta v_{total}$ has a maximum of
$\sim$ 6 km s$^{-1}$ (FWHM $\sim$ 14 km s$^{-1}$) near the O5 star
and decreases in the regions away from the O5 star. In terms of
mean velocity and velocity dispersion, the molecular gas near the
HII region is clearly disturbed. Besides the feature near the HII
region, the velocity dispersion along the \textit{sharp dust
ridge} is larger and has a correspondent extension (NE$-$SW),
which suggests that the molecular cloud along the \textit{sharp
dust ridge} is more turbulent (see also Sec. 4.3.1). This enhanced
turbulent motion supports our separation of the polarized emission
into component "o" and "x" in Sec. 4.2.2. These two polarized
components are most likely tracing different physical
environments.

The PV plots cut at various PAs at the position of the O5 star and
cut along the extension in the NE and SW direction on the 2nd
moment image (white segments on the lower panel of Fig. 5) are
shown in Fig. 6. The strongest emission is at $v_{sys}$ with an
extension of 18$\arcsec$, which suggests that the majority of the
gas is quiescent. Besides the quiescent gas, a ring-like
structure, indicated as red-dashed ellipses in Fig. 6, can be seen
clearly, especially at the PA of 60$\degr$ to 100$\degr$. Both an
infalling motion (e.g. Ho \& Young 1996) and an expansion can
produce a ring-like structure in the PV plots. In an infalling
motion, the expected free-fall velocity is $\sim$ 5 km s$^{-1}$
for a central mass of 50 M$_{\sun}$ at a distance of 2$\arcsec$
from the central star. This is smaller than the value measured in
the ring-like structure in G5.89. This C$^{17}$O 3-2 ring-like
structure in the PV plots is therefore more likely tracing the
expansion along with the HII region because of its high velocity
($\pm$10 km s$^{-1}$) and its dimension (2$\arcsec$ in radius).
However, the ring structure is not complete. This may be because
the material surrounding the HII region is not homogeneously
distributed, or the HII region is not completely surrounded by the
molecular gas.

Besides the expansion motion along with the HII region, there are
higher velocity components extending up to 30 km s$^{-1}$
(red-shifted) and $-$5 km s$^{-1}$ (blue-shifted) (Fig. 6). The
high velocity structure extending from the position of
2.5$\arcsec$ to the velocity of $\sim$30 km s$^{-1}$ is clearly
seen in the PV cuts at PA of 0$\degr$ to 40$\degr$ (indicated as
cyan arcs in Fig. 6). These high velocity components are probably
due to the sweeping motion of the molecular outflows in G5.89,
because there is no other likely energy source which can move the
material to such a high velocity. From the PV plots at the
position of the O5 star at various PAs and the PV plot cut along
the \textit{sharp dust ridge}, we conclude that the molecular
cloud is most likely both expanding along with the HII region and
being swept-up by the molecular outflows, all in addition to the
bulk of the quiescent gas.

The examination of the spectra at various positions also helps to
analyze the kinematics in G5.89. The spectra (Fig. 7) near the HII
ring (positions $C$, $D$, $F$, $G$ and $H$) have broad
line-widths. Furthermore, the spectra are not Gaussian-like, or
with distinct components at high velocities ($\pm$ 10 km
s$^{-1}$). At position $F$ and $H$, both spectra show a strong
peak at v$_{sys}$. The high velocity wing at the position $F$ is
red-shifted, and it is blue-shifted at position H. This is
consistent with the NS molecular outflow. The molecular gas near
the positions $E$ and $I$ is more quiescent because of its narrow
linewidth. The spectrum taken at position $I$ has a FWHM of 4 km
s$^{-1}$ and a peak intensity at $\sim$ 7 km s$^{-1}$. Comparing
with the spectra at other positions, the cloud around the position
$I$ is relatively quiescent and unaffected by the HII region or
the outflows. This cloud in the south near the position $I$ may be
a more independent component which is further separated along the
line of sight. We conclude that the C$^{17}$O 3-2 spectra
demonstrate that the kinematics and morphology have been strongly
affected by the expansion of the HII region. The nearly circular
structure in the PV plots, and in the channel maps near the
systemic velocity, as well as the spectra, suggest that a
significant part of the mass has been pushed by the HII region. An
impact from the molecular outflow can also be seen in the PV plots
and spectra.

\subsection{Estimate of the B field strength}

The B field strength projected in the plane of sky (B$_{\bot}$)
can be estimated by means of the Chandrasekhar-Fermi (CF) method
(Chandrasekhar \& Fermi 1953; Falceta-Gon\c{c}alves, Lazarian, \&
Kowal 2008). In general, the CF method can be applied to both dust
and line polarization measurements. We apply the CF method only to the
dust continuum polarization, because there is no line polarization
detected in this paper. Although the dust grains can also be
mechanically aligned, the existing observational evidence in NGC
1333 IRAS4A (Girart et al. 2006) demonstrates that the dust grains
can align with the B field in the low mass star formation regions.
Here, we assume that the dust grains also align with the B field
in G5.89.

The strength of B$_{\bot}$ can be calculated from:
\begin{equation}\label{1}
    B_{\bot} = Q \sqrt{4\pi\bar{\rho}}\frac{\delta v_{los, A}}{\delta\phi}
    = 63 \sqrt{n_{H_2}}\frac{\delta v_{los, A}}{\delta \phi}
\end{equation}

Here, B$_{\bot}$ is in the unit of mG. The term Q is a
dimensionless parameter smaller than 1. Q is $\sim$0.5 (Ostriker,
Stone \& Gammie 2001), depending on the inhomogeneities
within the cloud,
the anisotropies of the velocity perturbations, the observational
resolution and the differential averaging along the line of sight.
The term $\bar{\rho}$ is the mean density. $\delta \phi$ is the
dispersion of the polarization angles in units of degree. $\delta
v_{los, A}$ is the velocity dispersion along the line of sight in
units of km s$^{-1}$, which is associated with the Alfv\'{e}nic
motion. $n_{H{2}}$ is the number density of H$_{2}$ molecules in
units of 10$^{7}$ cm$^{-3}$. It has been shown numerically that
the CF method is a good approximation for $\delta \phi <
25^{\degr}$ (Ostriker, Stone, \& Gammie (2001)).

$\delta v_{los, A}$ is estimated from $\delta v_{total}$ in the
2nd moment image (lower panel in Fig. 5). $\delta v_{total}$
contains the information of the dispersions caused by the
Alfv\'{e}nic turbulent motion ($\delta \textit{v}_{los, A}$) and
the dispersions caused by the HII expansion and outflow motions
($\delta \textit{v}_{bulk}$). The relation of these three
components is:

\begin{displaymath}\label{1}
    \delta v_{total} \sim \sqrt{\delta v_{los, A}^{2}+\delta v_{bulk}^{2}}
\end{displaymath}

Here, we neglect the minor contributions from the thermal Doppler
motions. The measured $\delta v_{total}$ at the positions of
detected polarization are listed in Table 1. $\delta v_{total}$ is
in the range of 1 to 6 km s$^{-1}$. However, the molecular gas
near the HII region is clearly disturbed by both the HII expansion
and the molecular outflows (see Sec. 4.3.1 and 5.2). Therefore,
the detected $\delta v_{total}$ at these positions is dominated by
the bulk motion. Since $\delta v_{total}$ in the relatively
quiescent regions is more likely tracing the Alfv\'{e}nic motion
only, we adopt the minimum value $\delta v_{total}$ of 1 km
s$^{-1}$ at these positions for $\delta v_{los, A}$ in order to
derive B$_{\bot}$.

The term $n_{H_{2}}$ is $\sim$3$\times$10$^{6}$ (cm$^{-3}$),
estimated from the averaged $n_{H_{2}}$ from the 870$\mu$m dust
continuum and the C$^{17}$O 3-2 line emission (Sec. 4.1.2 and Sec.
4.3.2). $\delta\phi$ in Eq. 4 can be extracted from the observed
standard deviation of the PAs $\delta \phi_{obs}$. $\delta
\phi_{obs}$ contains both the observational uncertainty
$\sigma_{\phi,obs}$ and $\delta \phi$. The relation is:
$\delta\phi_{obs}^{2}$ = $\delta\phi^{2} + \sigma_{\phi,obs}^{2}$.
Since the polarization in G5.89 results probably from two
different systems (discussed in Sec. 4.2.1 and 4.2.2), it is more
reasonable to separate these two groups when deriving
$\delta\phi$. The derived $\sigma_{\phi,obs}$, $\delta\phi_{obs}$
and $\delta\phi$ are 3$\degr$ and 11$\degr$ and $\sim$11$\degr$
for component "o", and 2$\degr$, 18$\degr$ and $\sim$18$\degr$ for
component "x", respectively. By using Eq. (4), the derived
B$_{\bot}$ is 3mG and 2mG for component "o" and "x", respectively.

The estimated B$_{\bot}$ is highly uncertain. Due to the bulk
motions, it is difficult to extract the $\delta v_{los, A}$
component from the observed $\delta v_{total}$. The uncertainty
introduced from $\delta v_{los, A}$ is within a factor of 6. Of
course, the grouping of "o" and "x" components of the
polarization, as motivated in Sec. 4.2.1, 4.2.2 and 5.2, is not a
unique interpretation. If $\delta \phi$ is calculated without
grouping, a more complex model of the larger scale B field
morphology is needed to calculate the deviation due to the
Alfv\'{e}nic motion. More observations with sufficient uv coverage
are required to establish such a model. Based on the standard
deviation $\delta \phi$ of 39$\degr$ from the detected
polarization without subtracting the larger scale B field and
without grouping, the calculated lower limit of B$_{\bot}$ is
$\sim$1mG. Therefore, the estimated B$_{\bot}$ from the grouping
of component "o" and "x" seems reasonable. The value is comparable
to the ones estimated via the CF method in other massive star
formation regions with an angular resolution of a few arcseconds:
$\sim$ 1mG in DR 21(OH) (Lai et al. 2003) and $\sim$1.7 mG in
G30.79 FIR 10 (Cortes \& Crutcher 2006). Moreover, B$_{\bot}$ is
similar to B$_{\parallel}$ measured from the Zeeman pairs of the
OH masers by Stark et al. (2007), ranging from $-$2 to 2 mG.
Although B$_{\parallel}$ measured from OH masers is most likely
tracing special physical conditions, such as shocks or dense
regions, it is the only direct measurement of B$_{\parallel}$ in
G5.89, and hence, is of interest to compare. Assuming B$_{\bot}$
and B$_{\parallel}$ have the same strengths of 2mG, the total B
field strength in G5.89 is $\sim$ 3 mG.

\subsection{Collapsing cloud or not?}
The mass to flux ratio $\lambda$, a crucial parameter for the
magnetic support/ambipolar diffusion model, can be calculated
from: $\lambda$ = 7.6 $\times$ 10$^{-21}$ $\frac{N_{H_{2}}}{B}$
(Mouschovias \& Spitzer 1976; Nakano \& Nakamura 1978). $N_{H_2}$
is in cm$^{-2}$. $B$ is in $\mu$G. In the case of $\lambda$ $<$ 1,
the cloud is in a subcritical stage and magnetically supported. In
the case of $\lambda$ $>$ 1, the cloud is in a collapsing stage.

Since there is no observation of the B field strength as a
function of position in the entire cloud, we assume that the B
field is uniform with the strength of 3 mG in the entire cloud
when $\lambda$ is calculated. For consistency, when comparing with
the kinetic pressure in the next section, $N_{H_{2}}$ is derived
from the C$^{17}$O 3-2 emission (section 4.3.2). The derived
$\lambda$ in G5.89 is $>$1 in most parts of the molecular cloud,
as shown in the upper panel of Fig. 8. If the statistical
geometrical correction factor of $\frac{1}{3}$ is considered
(Crutcher 2004), the $\lambda_{corr}$ in the \textit{sharp dust
ridge} is still close to 1, whereas at the positions of the
component "o" and the outer part, it is much smaller than 1. This
suggests that G5.89 is probably in a supercritical phase near the
HII region and in a subcritical phase in the outer part of the
dust core.

This conclusion is based on the assumption that the B field in the
entire cloud is uniform with a strength of 3 mG. This assumption
seems to be crucial at first glance. However, the derived
$\lambda$ increases from 0.1 to 2.5 toward the UCHII region, which
is due to the high contrast of the column density across the
cloud. Unless the actual B field strength differs by a factor of
25 across the region and compensates for the contrast in the
column density, such a variation of $\lambda$ in G5.89 is indeed
possible.

\subsection{Compressed field?}
The coincident location of the detected polarization of component
"x" and the \textit{sharp dust ridge} is quite interesting. One
possible scenario is that the B field lines are compressed by the
shock front, i.e. HII expansion. In a magnetized large molecular
cloud with a B field traced by a component "o", and with a shock
sent out from the east of the narrow dust ridge, we expect a rapid
change of the polarization PA. This is similar to the results in
magnetohydrodynamic simulations by Krumholz et al. (2007). Because
of our limited angular resolution, polarization with a large
dispersion in the PAs over a small physical scale will be averaged
out within the synthesized beam. In our result, in fact, there is
a gap where polarized emission is not detected right next to the
\textit{sharp dust ridge}, and a series of OH masers are detected
in this gap. Note that the OH masers are most likely from the
shock front. From the discussion in Sec. 5.2, evidence for the
molecular cloud expanding with the HII region is found in the
molecular gas traced by the C$^{17}$O 3-2 emission. The 870$\mu$m
\textit{sharp dust ridge} can be explained by the swept-up
material along with the molecular gas from the HII expansion. In
this scenario, the component "o" is tracing the B field in the
pre-shock region, while the "x" component is tracing the
compressed field.

However, the swept-up flux density (summation of the flux density
of SMA-N, SMA-1 and SMA-2 reported in Hunter et al. 2008) is
$\sim$ 20\% of the total detected flux density in this paper. This
requires a huge amount of energy to sweep up the material with
this mass. Is the radiation pressure (P$_{rad}$) from the central
star large enough to overcome the kinetic pressure (P$_{kin}$) and
the B field pressure (P$_{B}$)? Here we compare these pressure
terms.

P$_{rad}$ can be calculated from the luminosity in G5.89 following
the equation: P$_{rad}$ $=$ $\frac{L}{cA}$, where $L$, $c$ and A
are the luminosity, speed of light and the area, respectively.
Since the G5.89 region is dense, most of the radiation is absorbed
and redistributed into the surrounding material. The total far
infrared luminosity of G5.89 is 3$\times$10$^{5}$L$_{\sun}$
(Emerson, Jennings, \& Moorwood 1973) and the radius of the HII
region at 2cm is $\sim$ 2$\arcsec$ (4000 AU). The energy density
and hence, the radiation pressure (P$_{rad}$) in the sphere with a
radius of 2$\arcsec$ is 8.5$\times$10$^{-7}$ (dyne cm$^{-2}$).

P$_{kin}$ is calculated by using the 0th moment ($MOM0$) and 2nd
moment ($MOM2$) images of the C$^{17}$O 3-2 line:
\begin{equation}\label{1}
    P_{kin} = \frac{1}{2} \rho \delta v_{total}^2 = 3.4 \times10^{-9} \times (MOM0) \times
    (MOM2)^2
 \end{equation}

where P$_{kin}$ is in dyne cm$^{-2}$, $\rho$ is the gas density in
g cm$^{-3}$ and $\delta v_{total}$ is the velocity dispersion in
cm s$^{-1}$, $MOM0$ is in units of K km s$^{-1}$, and $MOM2$ is in
units of km s$^{-1}$. $\rho$ is calculated following Sec. 4.3.2,
with the size of 0.13 pc for the molecular cloud along the line of
sight. The derived P$_{kin}$ image is shown in the middle and
lower panels of Fig. 8. P$_{kin}$ is in the range of $\sim$
1$\times$10$^{-9}$ to 1.4$\times$10$^{-6}$ dyne cm$^{-2}$.
P$_{kin}$ is calculated under the assumption that the length along
the line of sight is uniform in G5.89, which is the main bias in
the calculation. The estimated total B field strength is 3 mG,
thus the B-field pressure (P$_B$) is 3.6$\times$10$^{-7}$ (dyne
cm$^{-2}$). Although the upper limit of P$_{kin}$ is 1.6 times
larger than P$_{rad}$ at a radius of 2$\arcsec$, any variation of
the structure in the direction along the line of sight in G5.89 -
which is most likely the case - will affect the estimated
P$_{kin}$. Nevertheless, P$_{rad}$ is at the same order as
P$_{kin}$ and P$_{B}$ at the radius of 2$\arcsec$ around the O5
star. Therefore, in terms of pressure, the radiation from the
central star is likely sufficient to sweep up the material and
compress the B field lines along the narrow dust ridge.
$\lambda_{corr}$ close to 1 near the UCHII region suggests that
the B fields play a minor role as compared with the gravity.

The B field direction traced by component "x" is parallel to the
major axis of this sharp dust ridge, which is also seen in some
other star formation sites such as Cepheus A (Curran $\&$
Chrysostomou 2007) and DR 21(OH) (Lai et al. 2003). However, in
most of the cases, the detected B field direction is parallel to
the minor axis of the dust ridge, e.g. W51 e1/e2 cores (Lai et al.
2001) and G34.4+0.23 MM (Cortes et al. 2008), which agrees with
the ambipolar diffusion model. A possible explanation is that the
polarization is from the swept-up material, which interacts with
the original dense filament. Thus, the polarization here may
represent the swept-up field lines. This scenario is supported by
the energy density and also the morphology of the field lines in
the case of G5.89.

\subsection{Comparison with Other Star Formation Sites}
The detected B field structure in the G5.89 region is more
complicated than the B fields in other massive star formation
sites detected so far with interferometers. Both the compressed
field structure and the more organized larger scale B field are
detected in G5.89.

The B field lines vary smoothly in the cores at earlier star
formation stages, such as the W51 e1/e2 cores (Lai et al. 2001),
G34.4+0.23 MM (Cortes et al. 2008) and NGC 2024 FIR 5 (Lai et al.
2002). These cores are still in a collapsing stage (Ho et al. 1996
; Ramesh et al. 1997; Mezger et al. 2001). Among these
observations, the B fields inferred from the dust polarization
show an organized structure over the scale of 10$^{5}$ AU
(15$\arcsec$) at a distance of 7 kpc in the W51 e1/e2 cores and
also on the scale of 10$^{5}$ AU (35$\arcsec$) at a distance of
3.9 kpc in G34.4+0.23 MM. In contrast, the B fields in NGC 2024
FIR 5, which is closer at a distance of 415 pc, show an hourglass
morphology on a scale of 4$\times$10$^{3}$ AU (9$\arcsec$). Such
small scale structures would not be resolved in the current data
of W51 e1/e2 core and G34.4+0.23 MM due to the resolution effect.
Compared to these sources, G5.89 is more complicated with
polarization structures on both small (4$\times$10$^{3}$ AU) and
large (2$\times$10$^{4}$ AU) scales. However, higher angular
resolution polarization images of the cores in the earlier stages
are necessary in order to compare the B field morphology with the
later stages in the massive star formation process. At this
moment, we cannot conclude at which stage the B field structures
become more complex.

Currently, the best observational evidence supporting the
theoretical accretion model is the polarization observation of the
source NGC 1333 IRAS 4A (Girart, Rao \& Marrone 2006) carried out
with the SMA. The NGC 1333 IRAS 4A is a low mass star formation
site, at a distance of $\sim$300 pc. The detected pinched B-field
structure is at a scale of 2400 AU (8$\arcsec$). If NGC 1333 IRAS
4A were at a distance of 2kpc, we could barely resolve it at our
resolution of 3$\arcsec$. Higher angular resolution polarization
measurements are required to resolve the underlying structure in
G5.89.

\section{Conclusions and Summary}
High angular resolution (3$\arcsec$) studies at 870$\mu$m have
been made of the magnetic (B) field structures, the dust continuum
structures, and the kinematics of the molecular cloud around the
Ultra-Compact HII region G5.89-0.39. The goal is to analyze
the role of the B field in the massive star forming process. Here
is the summary of our results:

\begin{enumerate}
\item The gas mass (M$_{gas}$) is estimated from the dust
continuum and from the C$^{17}$O 3$-$2 emission line. The
continuum emission at 870 $\mu$m is detected with its total flux
density of 12.6$\pm$1.3 Jy. After removing the free-free emission
from the detected continuum, the flux density of the 870 $\mu$m
dust continuum is 7.7 Jy, which corresponds to M$_{gas}$
$\sim$300 M$_{\sun}$. M$_{gas}$ derived from the detected
C$^{17}$O 3-2 emission line is $\sim$100 M$_{\sun}$, which is 3
times smaller than the value derived from the dust continuum. The
discrepancy of M$_{gas}$ derived from the dust continuum and the
C$^{17}$O emission line is also seen in other UCHII regions, e.g.
Hofner et al. (2000). The lower values measured from C$^{17}$O
could be due to optical depth effects or abundance problems.

\item The linearly polarized 870 $\mu$m dust continuum emission is
detected and resolved. The dust polarization is not uniformly
distributed in the entire dust core. Most of the polarized
emission is located around the HII ring, and there is no
polarization detected in the southern half of the dust core except
at the very southern edges. The position angles (PAs) of the
polarization vary enormously but smoothly in a region of
2$\times$10$^{4}$ AU (10$\arcsec$), ranging from $-$60$\degr$ to
61$\degr$. Furthermore, the polarized emission is from organized
patches, and the distribution of the PAs can be separated into
two groups. We suggest that the polarization in G5.89
traces two different components. The polarization group "x", with
its PAs ranging from $-$60$\degr$ to $-$4$\degr$, is located at
the 870$\mu$m \textit{sharp dust ridge}. In contrast, the group
"o", with its PAs ranging from 33$\degr$ to 61$\degr$, is at the
periphery of the \textit{sharp dust ridge}. The inferred B field
direction from group "x" is parallel to the major axis of the
870$\mu$m dust ridge. One possible interpretation of the
polarization in group "x" is that it may represent swept-up B field
lines, while the group "o" traces more extended structures. In the
G5.89 region, both the large scale B field (group "o") and the
compressed B field (group "x") are detected.

\item By using the Chandrasekhar-Fermi method, the estimated
strength of B$_{\bot}$ from component "o" and from component "x"
is in between 2 to 3 mG, which is comparable to the Zeeman
splitting measurements of B$_{\parallel}$ from the OH masers,
ranging from $-$2 to 2 mG by Stark et al. (2007). The derived
lower limit of B$_{\bot}$ from the detected polarization without
grouping and without modeling larger scale B field is $\sim$1mG.
Assuming that B$_{\bot}$ and B$_{\parallel}$ have the same
strengths of 2 mG in the entire cloud, the derived $\lambda$
increases from 0.1 to 2.5 toward the UCHII region, which is due to
the high contrast of the column density across the cloud. Unless
the actual B field strength differs by a factor of 25 across the
region and compensates for the contrast in the column density,
such a variation of $\lambda$ in G5.89 is suggested. The corrected
mass to flux ratio ($\lambda_{corr}$) is closer to 1 near the HII
region and is much smaller than 1 in the outer parts of the dust
core. G5.89 is therefore most likely in a supercritical phase near
the HII region.

\item The kinematics of the molecular gas is analyzed using the
C$^{17}$O 3-2 emission line. From the analysis of the channel
maps, the position velocity plots and spectra, the molecular gas
in the G5.89 region is expanding along with the HII region, and it
is also possibly swept-up by the molecular outflows. Assuming the
size along the line of sight is uniform in G5.89, P$_{kin}$ is
in the range of $\sim$ 1$\times$10$^{-9}$ to 1.4$\times$10$^{-6}$
dyne cm$^{-2}$. The calculated radiation pressure (P$_{rad}$) at a
radius of 2$\arcsec$ and the B field pressure (P$_B$) with a field
strength of 3mG are 8.5$\times$10$^{-7}$ and 3.6$\times$10$^{-7}$
dyne cm$^{-2}$, respectively. Although the upper limit of
P$_{kin}$ is 1.6 times larger than P$_{rad}$ at a radius of
2$\arcsec$, any variation of the structure in the direction along
the line of sight in G5.89 - which is most likely the case - will
affect the estimated P$_{kin}$. Nevertheless, P$_{rad}$ is on the
same order as P$_{kin}$ and P$_{B}$ at the radius of 2$\arcsec$
around the O5 star. The scenario that the matter and B field in
the 870$\mu$m \textit{sharp dust ridge} have been swept-up is
supported in terms of the available pressure.
\end{enumerate}

G5.89 is in a more evolved stage as compared with the
corresponding structures of other sources in the collapsing phase.
The morphologies of the B field in the earlier stages of the
evolution show systematic or smoothly varying structures, e.g. on
the scale of 10$^{5}$ AU for W51 e1/e2 and G34.4+0.23 MM, and on
the scale of 4$\times$10$^{3}$ AU for NGC 2024 FIR 5. With the
high resolution and high sensitivity SMA data, we find that the B
field morphology in G5.89 is more complicated, being clearly
disturbed by the expansion of the HII region and the molecular
outflows. The large scale B field structure on the scale of
2$\times$10$^{4}$ AU in G5.89 can still be traced with dust
polarization. From the analysis of the C$^{17}$O 3-2 kinematics
and the comparison of the available energy density (pressure), we
propose that the B fields have been swept up and compressed.
Hence, the role of the B field evolves with the formation of the
massive star. The ensuing luminosity, pressure and outflows
overwhelm the existing B field structure.


\begin{figure}
\includegraphics[angle=0,scale=0.7]{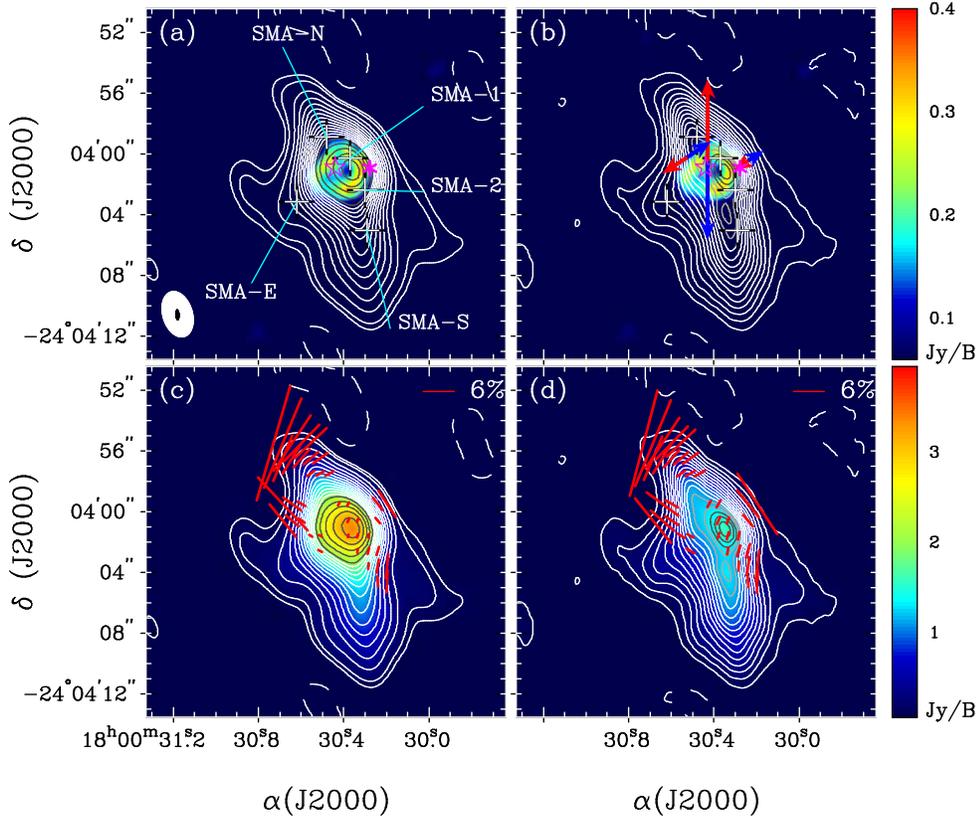}
\caption{\textit{(a)} The SMA 870 $\mu$m total continuum image
(contours) overlayed on the VLA 2cm continuum (free-free
continuum; color scale). The white contours represent the
continuum emission strength at 3, 5, 10, 15, 20, 25 ... 60 and 65
$\sigma$ levels, and the black contours in the center represent
70, 80, 90, 100 and 110 $\sigma$ levels, where {\rm $1\sigma$ is
30 mJy Beam$^{-1}$}. The star marks the O star detected by Feldt
et al. (2003). The asterisk marks the origin of the Br$\gamma$
outflow detected by Puga et al. (2006). The SMA and VLA
synthesized beams are shown as white and black ellipses at the
lower-left corner, respectively. The white "+" mark the positions
of the sub-mm peaks identified in Hunter et al. (2008).
\textit{(b)} The same as in (a), with the white contours
representing the SMA 870 $\mu$m dust continuum (after the
subtraction of the free-free continuum). The contours start from
and step in $3\sigma$, where {\rm $1\sigma$ is 30 mJy
Beam$^{-1}$}. The color wedge on the upper-right edge represents
the strength of the 2cm free-free continuum in the units of Jy
Beam $^{-1}$. The red and blue arrows indicate the axes of the
molecular outflows. The outflows in the N-S and NW-SW direction in
the west of O5 star are identified in Hunter et al. (2008). The
3rd outflow in the east of O5 star is identified in Tang et al.
(in prep.). \textit{(c)} The polarization (red segments) derived
by using image (a). The length of the red segment represents the
percentage of the polarized intensity. The 870 $\mu$m continuum is
shown both in white contours with the steps as in Fig. (a) and in
the color scale. \textit{(d)} The polarization (red segments)
derived by using image (b). The 870 $\mu$m dust continuum is again
shown both in white contours with the steps as in Fig. (b) and in
the color scale. The color wedge at the lower-right edge shows the
strength of the dust continuum in the units of Jy Beam$^{-1}$. In
\textit{(c)} and \textit{(d)}, the polarization plotted is above
3$\sigma_{I_p}$. } \label{}
\end{figure}

\begin{figure}
\includegraphics[angle=0,scale=0.6]{f2_a}
\includegraphics[angle=0,scale=0.6]{f2_b}
\includegraphics[angle=-90,scale=0.55]{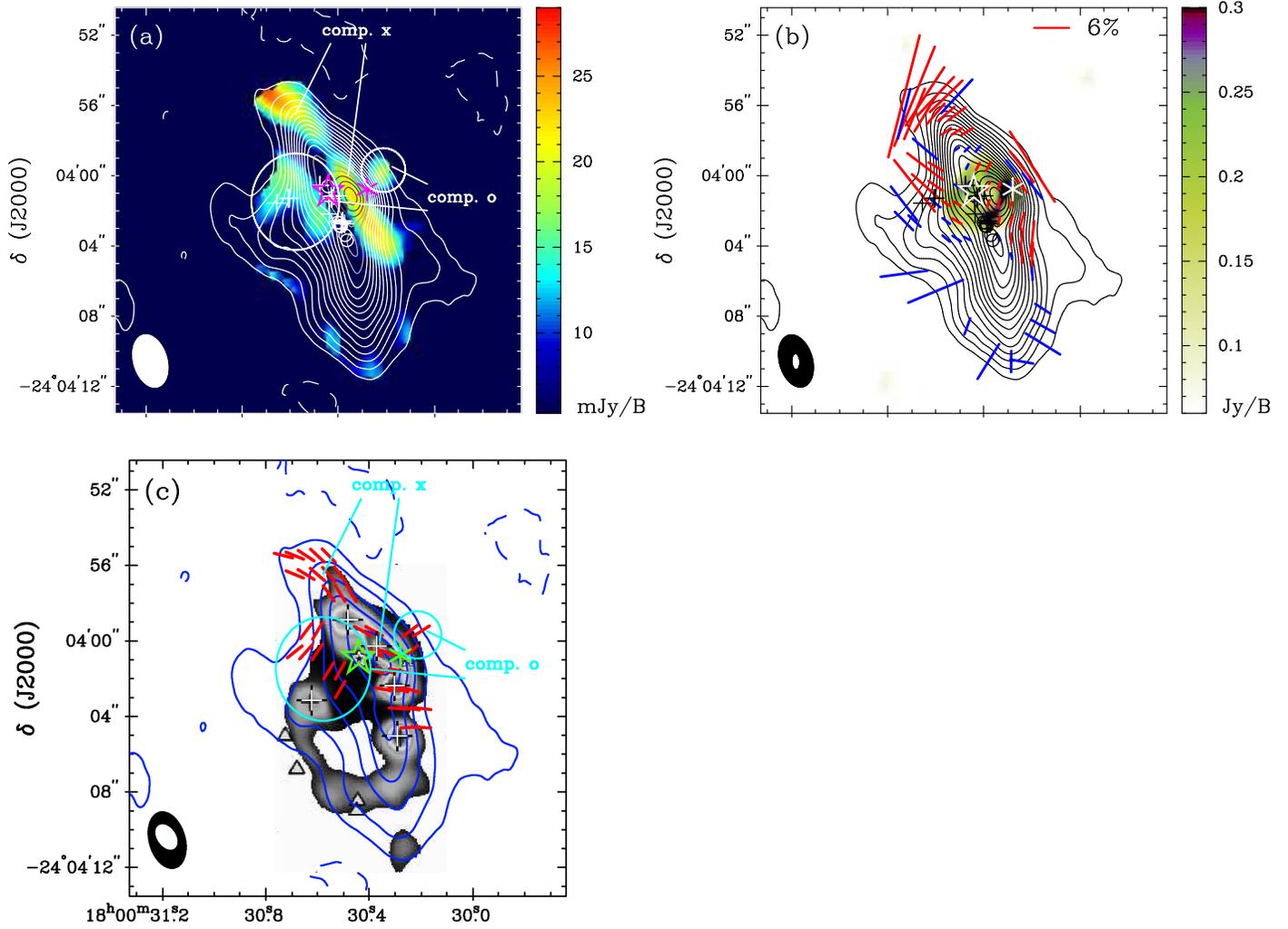}

\caption{(a) The 870$\mu$m dust continuum (white and grey
contours) overlaid on the polarized intensity (I$_{P}$) image
(color scale). The contours plotted are the same as in Fig. 1(b).
The color wedge shows the strength of polarization intensity in
units of mJy Beam$^{-1}$. The smallest white open circles and plus
signs mark the positions of the Zeeman pairs of the OH maser
(Stark et al. 2007) with different polarimetries. The other
symbols are the same as in Fig. 1. The larger solid white circles
mark component "o", defined in Sec. 4.2.2. (b) The polarization
(red and blue segments) overlaid on the 870 $\mu$m dust continuum
(black contours) and the 2cm free-free continuum emission (color
scale). In red and blue segments are polarization segments above
3$\sigma_{I_p}$ and between 2 to 3$\sigma_{I_p}$, respectively.
The contours, star, asterisk, circles and plus signs are all the
same as in (a). The color wedge shows the strength of the 2cm
free-free emission in units of Jy Beam$^{-1}$. The ellipses in the
lower-left corner are the synthesized beams of this paper, shown
in black, and of the 2cm free-free continuum image, shown in
white. (c) The inferred B field (red segments) overlaid on the 870
$\mu$m dust continuum (blue contours) in this paper and in Hunter
et al. (2008) (grey scale). The ellipses in the lower-left corner
are the synthesized beams of this work, shown in black, and of
Hunter et al. (2008), shown in white. The white crosses mark the
sub-mm sources detected by Hunter et al. (2008). The triangles
mark the positions of the H$_{2}$ knots identified in Puga et al.
(2006).} \label{pol}
\end{figure}

\begin{figure}
\includegraphics[angle=0,scale=0.6]{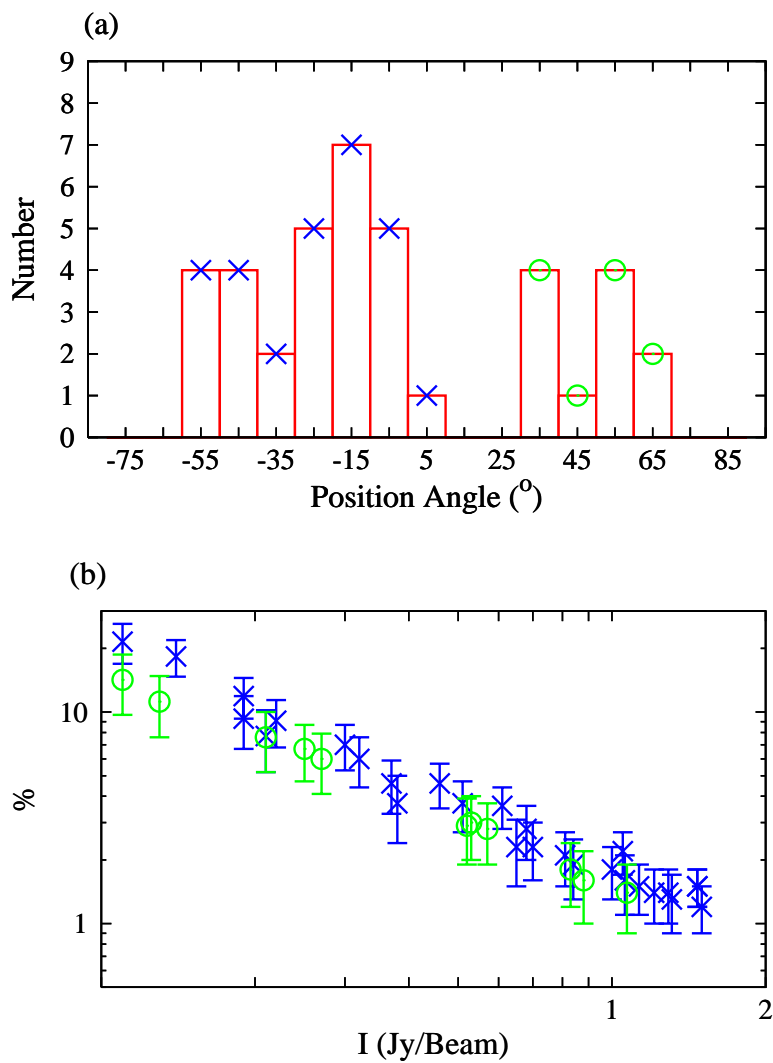} \caption{(a)
Distribution of the PAs (defined in the range $-$90$\degr$ to
90$\degr$) of the polarization in G5.89. (b) Total intensity (I)
versus percentage (\%) of the polarized flux density. In both (a)
and (b) panels, the statistics are from the detected polarization
segments above 3 $\sigma_{I_p}$ confidence level. The "cross"
represents the component "x", which is associated with the sharp
dust ridge. The "circle" represents the component "o", which is
associated with the extended structure.} \label{vec_dist}
\end{figure}

\begin{figure}
\includegraphics[angle=-90,scale=0.7]{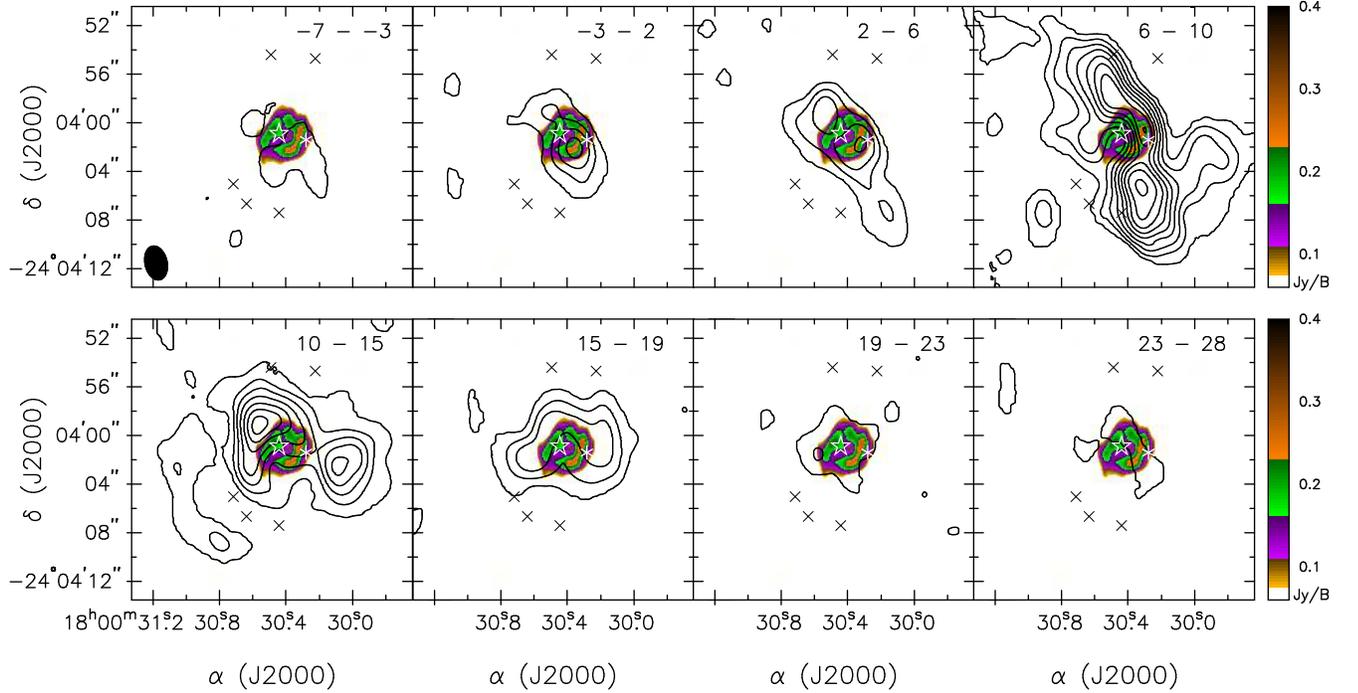}
\caption{The C$^{17}$O 3-2 line emission (black contours)
integrated in 8 different velocity ranges, as indicated in each
panel in units of km s$^{-1}$. The color scale images represent
the VLA 2cm free-free continuum emission with its strength shown
in the wedge in units of Jy Beam$^{-1}$. To show the fainter
component in the higher velocity panels, the black contours
represent the strengths at 2,6,10,15,21,27,33,40,48 $\sigma$,
where 1 $\sigma$ is 0.4 Jy Beam$^{-1}$ km s$^{-1}$. The black
crosses mark the positions of the H$_{2}$ knots detected by Puga
et al. (2006). The other symbols are the same as in Fig. 1. In the
panels from $-$7 to $-$3 and 23 to 28 km s$^{-1}$, the C$^{17}$O
dense cloud has an extension similar to the dust ridge detected in
the 0.8$\arcsec$ dust continuum image (Fig. 2). The peak positions
of the high velocity components do not coincide with the location
of the O5 star.}
\end{figure}
\begin{figure}
\includegraphics[angle=0,scale=0.6]{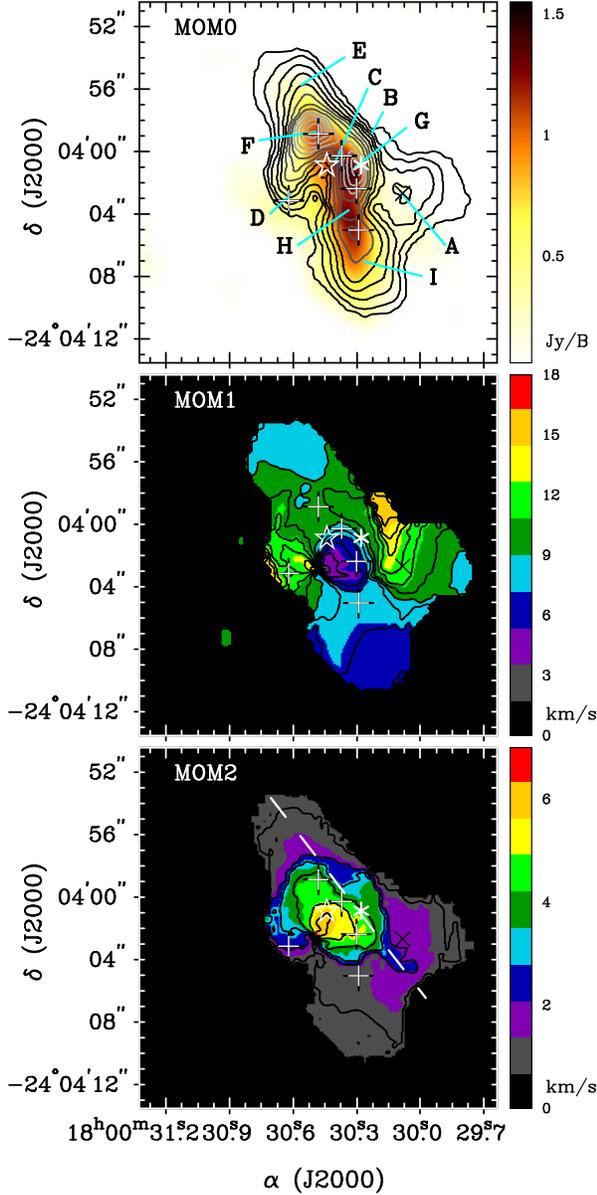}
\caption{\textsl{Upper Panel:} The total integrated intensity (0th
moment) image of the C$^{17}$O 3-2 emission line (black contours)
overlaid on the dust continuum at 870 $\mu$m (color scale). The
contours represent the emission strength starting from and
stepping in 3 Jy Beam$^{-1}$ km s$^{-1}$. The letters mark the
positions where the spectra were taken in Fig. 7. The black cross
marks the peak emission at position A. The color wedge shows the
strength of the 870 $\mu$m dust continuum in units of Jy
Beam$^{-1}$. The black contours start from and step in 3 $\sigma$,
where 1$\sigma$ is 1 Jy Beam$^{-1}$ km s$^{-1}$. \textsl{Middle
Panel:} The intensity weighted velocity image (1st moment) of the
C$^{17}$O 3-2 emission line. The contours start from and step in 1
km s$^{-1}$. The wedge is in units of km s$^{-1}$. \textsl{Lower
Panel:} The intensity weighted velocity dispersion ($\delta
v_{total}$) image (2nd moment) of the C$^{17}$O 3-2 emission line.
The contours start from and step in 1 km s$^{-1}$. The wedge is in
units of km s$^{-1}$. The white segments mark the cut of the
position velocity diagram shown in Fig. 6($j$). The rest of the
symbols are the same as in Fig. 1.}
\end{figure}
\begin{figure}
\includegraphics[scale=0.8,angle=-90]{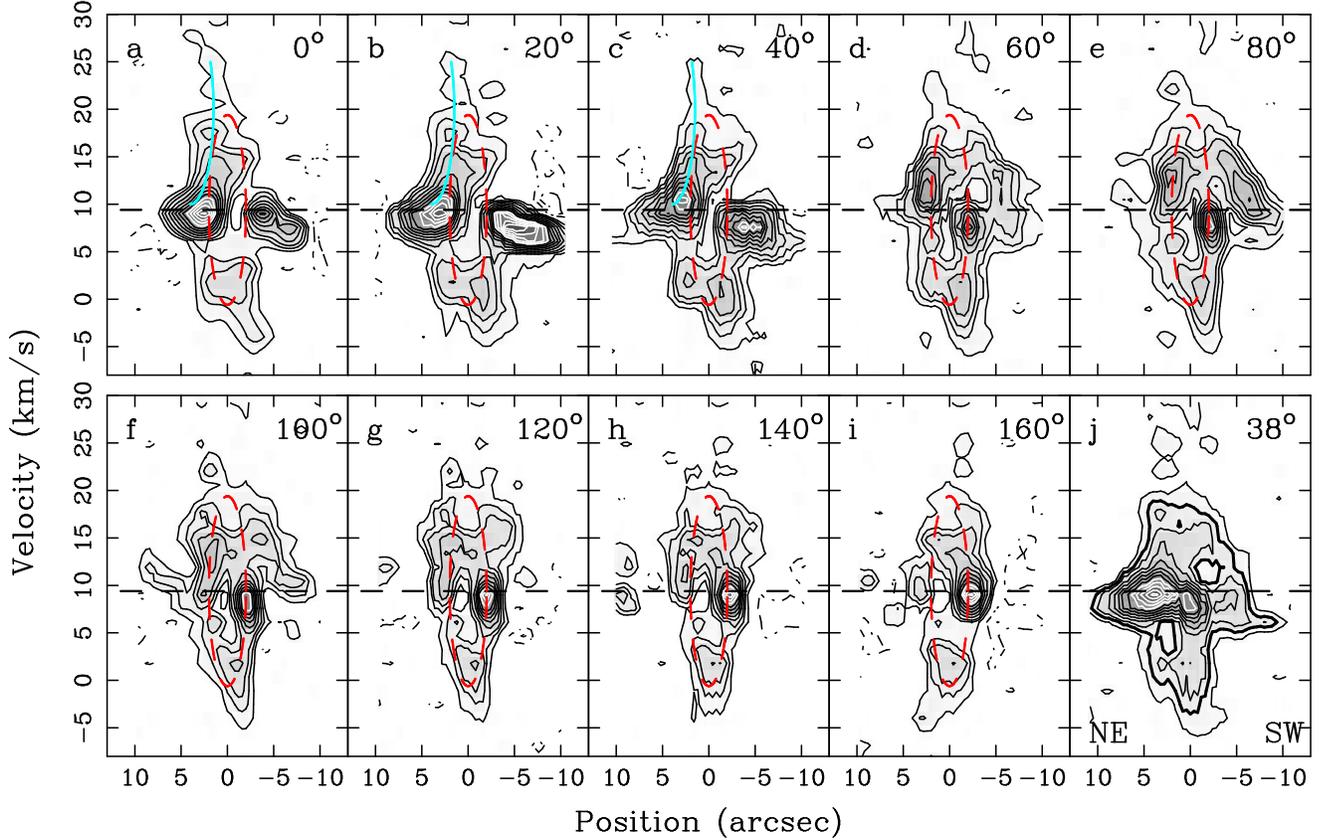}
\caption{The position-velocity (PV) diagram cuts at the position
of the O5 star (panel $a$ to $i$) at 9 different position angles,
as indicated at the upper-right corner in each panel. Panel $j$ is
the PV diagram along the white segments indicated in the lower
panel of Fig. 5. To show the fainter emission at outer regions,
the black contours are plotted in 2, 4, 6, 9, 12, 15, 18 and 21
$\times$ $\sigma$ and white contours in 24, 27, 30, 33, 36
$\times$ $\sigma$, where 1$\sigma=$0.16 Jy Beam$^{-1}$. The thick
contour in panel $j$ represents the emission strength of
4$\sigma$. The dashed segments indicate the v$_{sys}$ of 9.4 km
s$^{-1}$. The red dashed ellipse marks the ring-like structure
with a radius of 2$\arcsec$ and 10 km s$^{-1}$ and centered at the
position of the O5 star and v$_{sys}$ of 9.4 km s$^{-1}$. The
majority of the gas is quiescent with a velocity close to
$v_{sys}$. In panel $j$, a loop-like structure is seen at
positions of $\pm$ 2$\arcsec$. A clump at a v$_{lsr}$ of 0 km
s$^{-1}$ is detected in all the PV cuts at different PA (panel $a$
to $i$). At PA of 0$\degr$ to 40$\degr$, the high velocity
structure extending from the position of 2.5$\arcsec$ is clearly
seen, as indicated by the blue arc.}
\end{figure}

\begin{figure}
\includegraphics[angle=0,scale=0.9]{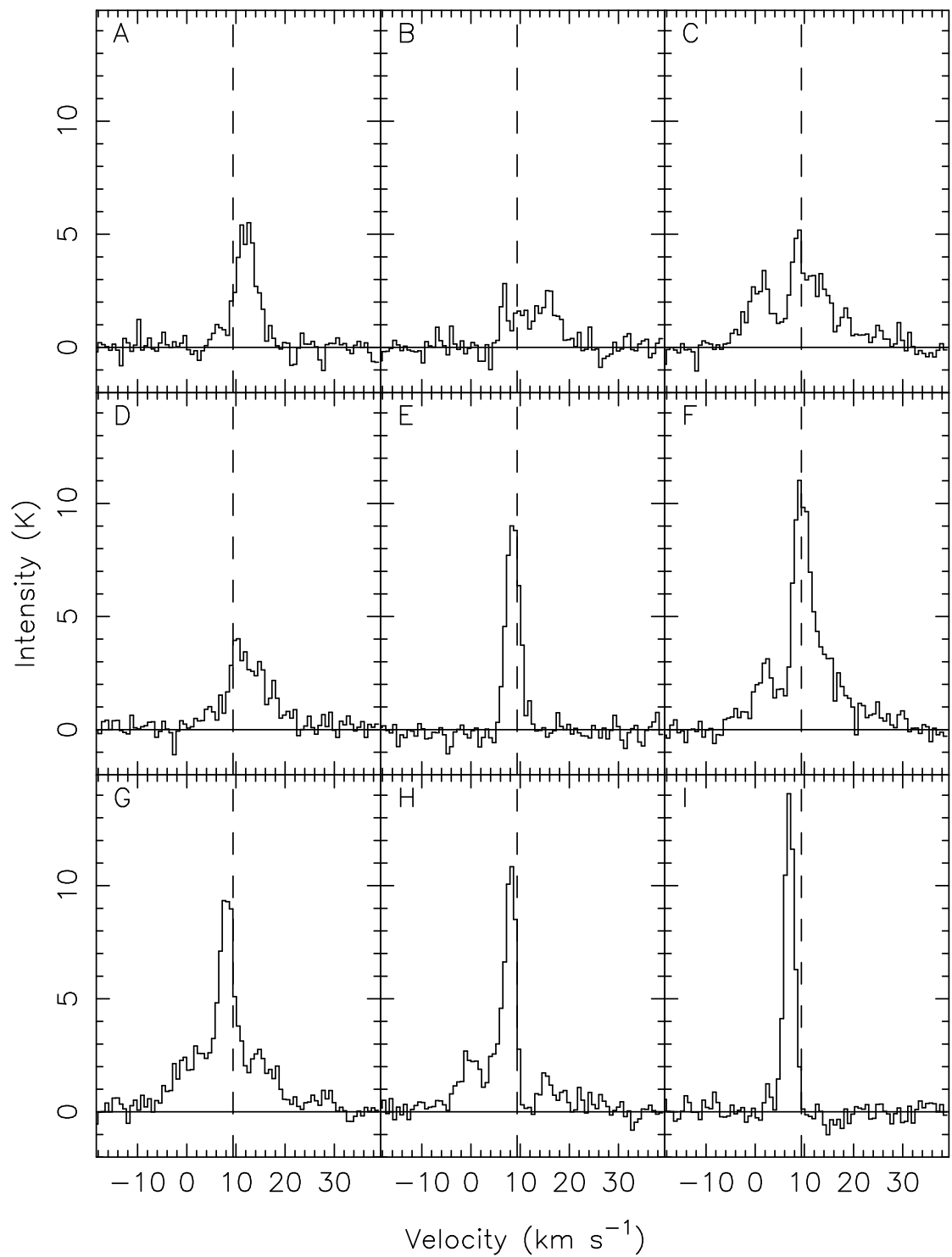}
\caption{The spectra of the C$^{17}$O 3-2 line at various
positions as indicated in Fig. 5. The solid and dashed segments
mark the intensity of 0 K and v$_{sys}$ of 9.4 km s$^{-1}$,
respectively. Note that the typical HII region expands at a
velocity of $\pm$ 10 km s$^{-1}$. At position E and I, the spectra
are narrowest.}
\end{figure}
\begin{figure}
\includegraphics[angle=0,scale=0.6]{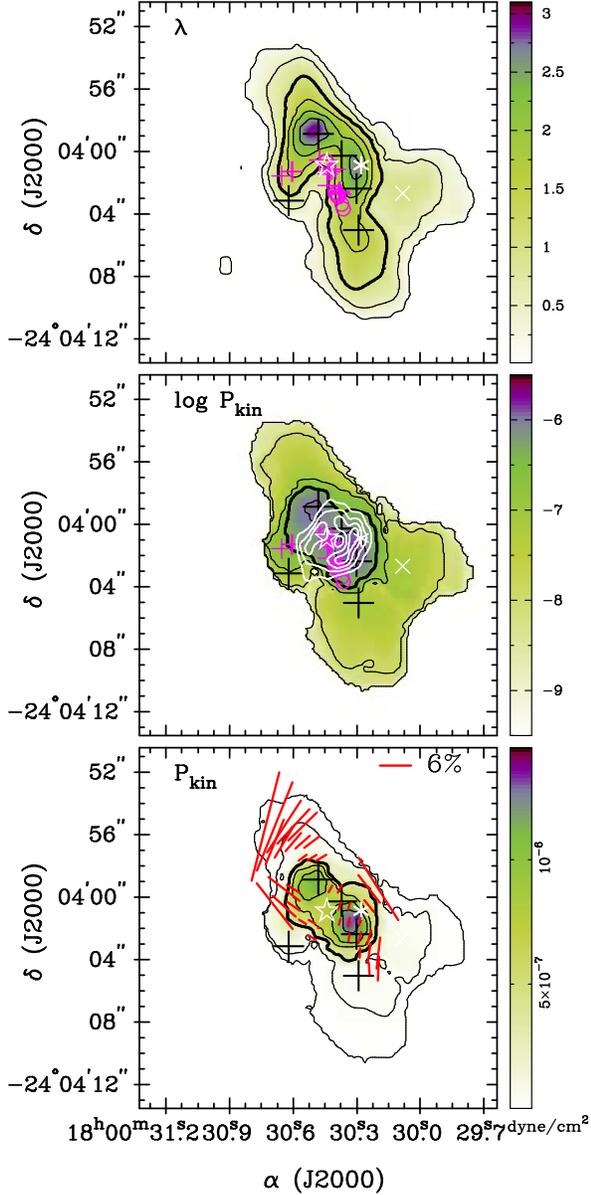}
\caption{\textit{Upper panel:} Image of the mass to flux ratio
$\lambda$ in both contours and color scale. $\lambda$ is derived
using the C$^{17}$O 3-2 emission line assuming a uniform total B
field strength of 3 mG, which corresponds to a B field pressure
(P$_{B}$) of 3.6$\times$10$^{-7}$ dyne cm$^{-2}$. The thick black
contour marks $\lambda$=1. The thin black contours mark
$\lambda$=0.1, 0.5, 1.5, 2, 2.5. The strength of $\lambda$ in
color scale is indicated in the wedge. The other symbols are the
same as in Fig. 2. $\textit{Middle panel:}$ Image of the derived
kinetic pressure (P$_{kin}$) in log scale in both black contours
and color scale. The thick black contour marks $log$ $P_{kin}$ $=$
$-$6.45, which is equal to P$_{B}$=3.6$\times$10$^{-7}$ dyne
cm$^{-2}$. The thin black contours mark $log$ P$_{kin}$ = $-$6,
$-$7, $-$8, and $-$9. The white contours represent the 2cm
free-free emission starting from and stepping in 72 mJy
Beam$^{-1}$. The wedge is in the units of $log$ dyne cm$^{-2}$.
$\textit{Lower panel:}$ Image of P$_{kin}$ in linear scale in both
black contours and color scale. P$_{kin}$ is in the unit of dyne
cm$^{-2}$. The thick black contour marks P$_{kin}$ $=$
3.6$\times$10$^{-7}$ dyne cm$^{-2}$. The thin black contours mark
P$_{kin}$ = (0.3, 3, 60, 90, 120)$\times$10$^{-8}$. The red
segments represent the 870 $\mu$m polarization above
3$\sigma_{I_p}$, detected in this paper.}
\end{figure}

\include{tab1}

\end{document}

%% file: tab1.tex
\begin{deluxetable}{ccccccl}

\tablecaption{SMA dust polarization at 870 $\mu m$}
\tablewidth{0pt} \tablehead{ \colhead{$\triangle$x} &
\colhead{$\triangle$y} & \colhead{I$_{p}$} & \colhead{\%} &
\colhead{PA} & \colhead{$\delta v_{total}$}& \colhead{group}}

\startdata
5.4 &   5   &   24  &   21.5    $\pm$   4.6 &   -14 $\pm$   6   &   1.0     &   x   \\
4.8 &   5   &   26  &   18.3    $\pm$   3.6 &   -22 $\pm$   5   &   1.2     &   x   \\
4.2 &   5   &   23  &   11.9    $\pm$   2.6 &   -33 $\pm$   6   &   1.2     &   x   \\
3.6 &   5   &   20  &   9.1 $\pm$   2.3 &   -43 $\pm$   7   &   1.2     &   x   \\
3   &   5   &   16  &   7.7 $\pm$   2.5 &   -46 $\pm$   9   &   1.2     &   x   \\
4.8 &   4   &   18  &   9.3 $\pm$   2.6 &   -22 $\pm$   8   &   1.0     &   x   \\
4.2 &   4   &   21  &   7   $\pm$   1.7 &   -31 $\pm$   7   &   1.5     &   x   \\
3.6 &   4   &   21  &   4.6 $\pm$   1.1 &   -41 $\pm$   7   &   1.5     &   x   \\
3   &   4   &   19  &   3.7 $\pm$   1   &   -49 $\pm$   8   &   1.5     &   x   \\
2.4 &   4   &   14  &   3.7 $\pm$   1.3 &   -53 $\pm$   10  &   1.4     &   x   \\
3   &   3   &   16  &   1.9 $\pm$   0.6 &   -54 $\pm$   9   &   2.5     &   x   \\
2.4 &   3   &   17  &   2.1 $\pm$   0.6 &   -59 $\pm$   8   &   2.2     &   x   \\
1.8 &   3   &   15  &   2.3 $\pm$   0.8 &   -60 $\pm$   10  &   2.0     &   x   \\
4.2 &   1   &   16  &   7.6 $\pm$   2.4 &   53  $\pm$   9   &   3.4     &   o   \\
3.6 &   1   &   16  &   2.8 $\pm$   0.9 &   60  $\pm$   9   &   4.3     &   o   \\
1.2 &   1   &   17  &   1.4 $\pm$   0.4 &   -29 $\pm$   8   &   4.0     &   x   \\
0.6 &   1   &   17  &   1.6 $\pm$   0.5 &   -29 $\pm$   8   &   3.3     &   x   \\
-1.2    &   1   &   17  &   6.7 $\pm$   2   &   38  $\pm$   9   &   3.8     &   o   \\
-1.8    &   1   &   16  &   14.2    $\pm$   4.5 &   33  $\pm$   9   &   1.8     &   o   \\
4.8 &   0   &   15  &   11.2    $\pm$   3.6 &   38  $\pm$   9   &   2.7     &   o   \\
4.2 &   0   &   16  &   6   $\pm$   1.9 &   48  $\pm$   9   &   3.6     &   o   \\
3.6 &   0   &   16  &   3   $\pm$   1   &   56  $\pm$   9   &   4.0     &   o   \\
0.6 &   0   &   22  &   1.5 $\pm$   0.3 &   -24 $\pm$   7   &   4.6     &   x   \\
0   &   0   &   17  &   1.3 $\pm$   0.4 &   -18 $\pm$   9   &   4.5     &   x   \\
-1.2    &   0   &   15  &   2.9 $\pm$   1   &   39  $\pm$   10  &   3.9     &   o   \\
3   &   -1  &   14  &   1.6 $\pm$   0.6 &   58  $\pm$   10  &   4.5     &   o   \\
2.4 &   -1  &   15  &   1.4 $\pm$   0.5 &   61  $\pm$   9   &   5.8     &   o   \\
0.6 &   -1  &   18  &   1.2 $\pm$   0.3 &   -16 $\pm$   8   &   5.3     &   x   \\
0   &   -1  &   22  &   1.5 $\pm$   0.3 &   -15 $\pm$   6   &   5.0     &   x   \\
-0.6    &   -1  &   17  &   1.5 $\pm$   0.4 &   -12 $\pm$   8   &   4.8     &   x   \\
2.4 &   -2  &   15  &   1.8 $\pm$   0.6 &   59  $\pm$   9   &   5.9     &   o   \\
0   &   -2  &   18  &   1.4 $\pm$   0.4 &   -10 $\pm$   8   &   5.0     &   x   \\
-0.6    &   -2  &   23  &   2.2 $\pm$   0.5 &   -14 $\pm$   6   &   4.8     &   x   \\
-1.2    &   -2  &   19  &   2.8 $\pm$   0.8 &   -16 $\pm$   8   &   4.7     &   x   \\
-0.6    &   -3  &   18  &   1.8 $\pm$   0.5 &   -4  $\pm$   8   &   3.3     &   x   \\
-1.2    &   -3  &   22  &   3.6 $\pm$   0.8 &   -8  $\pm$   6   &   2.4     &   x   \\
-1.8    &   -3  &   19  &   6   $\pm$   1.6 &   -6  $\pm$   8   &   2.7     &   x   \\
-1.2    &   -4  &   16  &   2.3 $\pm$   0.7 &   4   $\pm$   9   &   1.4     &   x   \\
-1.8    &   -4  &   17  &   4.6 $\pm$   1.3 &   -3  $\pm$   8   &   1.4     &   x   \\

\enddata
\tablecomments{$\triangle$x \& $\triangle$y: offsets in arcsecond
from the coordinate (J2000): $\alpha=18:00:30.32$,
$\delta=-24:04:00.48$. $I_{p}$: the polarized intensity in mJy
Beam$^{-1}$. \%: polarization percentage, defined as the ratio of
I$_{p}$/I. PA: position angle from the north to the east in
degree. $\delta v_{total}$: total dispersion velocity (2nd moment)
measured in the C$^{17}$O 3-2 emission line in km s$^{-1}$. All
data listed are above 3$\sigma_{I_{p}}$.}
\end{deluxetable}

%% file: ms_v3.bbl
\begin{thebibliography}{}

\bibitem[]{} Acord, J. M., Churchwell, E., \& Wood, D. O. S. 1998, ApJ,
495, 107
\bibitem[]{} Cesaroni, R., Walmsley, C. M., Koempe, C., \&
Churchwell, E. 1991, A\&A, 252, 278

\bibitem[]{} Chandrasekhar, S., \& Fermi, E. 1953, ApJ, 118, 113
\bibitem[]{} Choi, M., Evans II, N., \& Jaffe, D. T. 1993, ApJ, 417, 624

\bibitem[]{} Churchwell 1997, ApJ, 479, L59



\bibitem[]{} Cortes, P. C., Crutcher, R. M., \& Watson, W. D. 2005,
628, 780
\bibitem[]{} Cortes, P., \& Crutcher, R. M. 2006, ApJ, 639, 965

\bibitem[]{} Cortes, P., Crutcher, R. M., \& Matthews, B. 2006,
ApJ, 650, 246

\bibitem[]{} Cortes, P., Crutcher, R. M., Shepherd, D. S., \&
Bronfman, L. 2008, ApJ, 676, 464

\bibitem[]{} Crutcher, R. M. 2004, ApSS, 292, 225

\bibitem[]{} Curran, R. L. \& Chrysostomou, A. 2007, MNRAS, 382,
699

\bibitem[]{} Deguchi, S. \& Watson, W. 1984, ApJ, 285, 126

\bibitem[]{} Draine, \& Weingartner 1996, ApJ, 470, 551

\bibitem[]{} Elmegreen, B. G., \& Scalo, J. 2004, ARA\&A, 42, 211

\bibitem[]{} Emerson, J. P., Jennings, R. E., \& Moorwood, A. F.
M. 1973, ApJ, 184, 401

\bibitem[]{} Falceta-Gon\c{c}alves, D., Lazarian, A., \& Kowal, G.
2008, ApJ, 679, 537

\bibitem[]{} Feldt, M., Stecklum, B., Henning, Th., Launhardt, R., \& Hayward, T.
L. 1999, A\&A, 346, 243

\bibitem[]{} Feldt, M., Puga, E., Lenzen, R., Henning, Th., Brandner, W., Stecklum, B.,
Lagrange, A.-M., Gendron, E., \& Rousset, G. 2003, ApJ, 599, L91

\bibitem[]{} Fiedler, R. A., \& Mouschovias, T. Ch. 1993, ApJ,
415, 680

\bibitem[]{} Fish, V. L, Reid, M. J., Argon, A. L., \& Zheng,
X.-W. 2005, ApJS, 160, 220

\bibitem[]{} Frerking, M. A., \& Langer, D. L., \& Wilson, W. W. 1982,
ApJ, 262, 590

\bibitem[]{} Galli, D., \& Shu, F. H. 1993, ApJ, 417, 243

\bibitem[]{} Girart, J. M., Crutcher, R. M. \& Rao, R. 1999, ApJ, 525, L109

\bibitem[]{} Girart, J. M., Rao, R., \& Marrone, D. P. 2006, Sci, 313, 812
\bibitem[]{} Goldreich, P., \& Kylafis, N. D. 1981, ApJ, 243, 75
\bibitem[]{} Gon\c{c}alves, J., Galli, D., \& Walmsley, M. 2005,
A\&A, 430, 979


\bibitem[]{} Harvey, P. M., \& Forveille, T. 1988, A\&A, 197, L19



\bibitem[]{} Ho, P. T. P., Moran, J. M., \& Lo, K. Y. 2004, ApJ,
616, 1

\bibitem[]{} Ho, P. T. P., \& Young, L. M. 1996, ApJ, 472, 742

\bibitem[]{} Hofner, P., \& Churchwell 1996, A\&AS, 120, 283

\bibitem[]{} Hofner, P., Wyrowski, F., Walmsley, C. M., \&
Churchwell, E. 2000, ApJ, 536, 393

\bibitem[]{} Hunter, T. R., Churchwell, E., Watson, C., Cox, P.,
Benford, D. J., \& Roelfsema, P. R. 2000, AJ, 119, 2711
\bibitem[]{} Hunter, T. R., Brogan, C. L., Indebetouw, R., \& Cyganowski, C.
J. 2008, 680, 127


\bibitem[]{} Kramer, C., Alves, J., Lada, C., Lada, E., Sievers,
A., Ungerechts, H., \& Walmsley, M. 1999, A\&A, 342, 257

\bibitem[]{} Krumholz, M., Stone, J. M., \& Gardiner, T. A. 2007,
ApJ, 671, 518
\bibitem[]{} Kurtz, S., Hofner, P., \& Alvarez, C. V. 2004, ApJS, 155, 149

\bibitem[]{} Kylafis, N. D. 1983, ApJ, 267, 137

\bibitem[]{} Lai, S.-P., Crutcher, R. M., Girart, J. M., \& Rao,
R. 2001, ApJ, 561, 864

\bibitem[]{} Lai, S.-P., Crutcher, R. M., Girart, J. M., \& Rao,
R. 2002, ApJ, 566, 925
\bibitem[]{} Lai, S.-P., Girart, J. M., \& Crutcher, R. M. 2003,
ApJ, 598, 392
\bibitem[]{} Lazarian, A. 2007, Journal of Quantitative
Spectroscopy \& Radiative Transfer, 106, 255

\bibitem[]{} Lazarian, A. \& Hoang, T. 2007, MNRAS, 378, 910


\bibitem[]{} Lis, D. C., Serabyn, E., Keene, Jocelyn, Dowell, C. D., Benford, D. J.,
Phillips, T. G., Hunter, T. R., Wang, N. 1998, 509, 299 Mac Low

\bibitem[]{} Mac Low, M.-M., \& Klessen, R. S. 2004, Rev. Mod. Phys., 76, 125

\bibitem[]{} Marrone, D. \& Rao, R. 2008, arXiv:0807.2255

\bibitem[mezger(1992)]{} Mezger, P. G., Sievers, A. W., Haslam, C.
G. T., Kreysa, E., Lemke, R., Mauersberger, R., \& Wilson, T. L.
1992, A\&A, 256, 631


\bibitem[]{} Mouschovias, T. Ch. 1976, ApJ, 207, 141

\bibitem[]{} Mouschovias, T. Ch., \& Spitzer, L. 1976, ApJ, 210,
326

\bibitem[]{} Mouschovias, T. Ch. \& Ciolek, G. E. 1999, in The
Origin of Stars and Planetary Systems, ed. C. J. Lada \& N. D.
Kylafis (Kluwer: Dordrecht), p. 305


\bibitem[]{} Nakano, T., \& Nakamura, T. 1978, PASJ, 30, 681


\bibitem[ostriker(2001)]{ostriker2001} Ostriker, E. C., Stone, J. M., \& Gammie, C.
F. 2001, ApJ, 546, 980


\bibitem[]{} Puga, E., Feldt, M., Alvarez, C., Henning, Th., Apai, D., Coarer, E. Le, Chalabaev, A.,
\& Stecklum, B. 2006, ApJ, 641, 373


\bibitem[]{} Ramesh, B., Bronfman, L., \& Deguchi, S. 1997, PASJ,
49, 307

\bibitem[]{} Rao, R., Crutcher, R. M., Plambeck, R. L., \& Wright,
M. C. H. 1998, ApJ, 502, L75
\bibitem[]{} Rohlfs, K., \& Wilson, T. L. 2004, Tools of Raido
Astronomy (4th ed; Berlin: Springer)
\bibitem[]{} Sault, R. J., Teuben, P. J., \& Wright, M. C. H.
1995, in ASP Conf. Ser. 77, Astronomical Data Analysis Software
and Systems IV, ed. R. A. Shaw, H. E. Payne, \& J. J. E. Hayes
(San Francisco: ASP), 433

\bibitem[]{} Sault, R. J., Hamaker, J. P., \& Bregman, J. D. 1996,
A\&AS, 117, 149

\bibitem[]{} Shu, F., Allen, A., Shang, H., Ostriker, E. C., \&
Li, Z.-Y. 1999, in The Origin of Stars and Planetary Systems, ed.
Charles J. Lada \& Nikolaos D. Kylafis, (Kluwer: Dordrecht), p.
193

\bibitem[Sollins(2004)]{sollins04} Sollins, P. K., Hunter, T. R., Battat, J., Beuther, H., Ho, P. T. P.,
Lim, J., Liu, S. Y., Ohashi, N., Sridharan, T. K., Su, Y. N.,
Zhao, J.-H., \& Zhang, Q. 2004, ApJ, 616, 35

\bibitem[]{} Stark, D. P., Goss, W. M., Churchwell, E. Fish, V. L., \& Hoffman, I. M. 2007, ApJ, 656, 943



\bibitem[]{} Watson, C., Churchwell, E., Zweibel, E. G., \& Crutcher, R. M. 2007, ApJ, 657, 318


\bibitem[]{} Wood, D. O. S., \& Churchwell, E. 1989, ApJS, 69, 831


\bibitem[]{} Zijlstra, A. A., Pottasch, S. R., Engels, D.,
Roelfsema, P. R., Hekkert, P. T., \& Umana, G. 1990, MNRAS, 246,
217

\end{thebibliography}
